\documentclass[aps,nofootinbib,superscriptaddress,preprintnumbers,12pt]{revtex4-1}
\pdfoutput=1  %needed for JHEP at least?
\usepackage[pdftex]{graphicx} %is this useful?
\usepackage{amsmath,amssymb,amsfonts,amsthm,epsf,dsfont,dcolumn,yfonts}  %omitted graphicx from here
\usepackage{latexsym}
\usepackage{hyperref} %colored hyperlinks
\usepackage{bm,mathrsfs,slashed} % generic pages setups
\usepackage{setspace} %just loading this corrects spacing in footnotes

\usepackage{color}
\newcommand{\be}{\begin{equation}}
\newcommand{\ee}{\end{equation}}
\newcommand{\bea}{\begin{eqnarray}}
\newcommand{\eea}{\end{eqnarray}}
\newcommand{\bwt}{\begin{widetext}}
\newcommand{\ewt}{\end{widetext}}

\newcommand{\nn}{\nonumber\\}

\newcommand{\dif}{{\rm d}}

\newcommand{\rmd}{{\rm d}}

  \def\p{\partial} 
  
 \newcommand{\tr}{\mathop{\rm Tr}} 
\def\expec#1{\langle #1 \rangle}

\newcommand{\cE}{{\mathcal{E}}}
\newcommand{\cF}{{\mathcal{F}}}

\newcommand{\cN}{{\mathcal{N}}}

\newcommand{\bR}{{\mathbf{R}}}
\newcommand{\bS}{{\mathbf{S}}}

\newcommand{\trFsq}{\tr F^2}%{{\mathcal O}_{F^{2}}}
\newcommand{\ups}{{\upsilon}}

\newcommand{\tv}{{\tilde{\upsilon}}}
\newcommand{\ta}{{\tilde{a}}}
\newcommand{\tx}{{\tilde{x}}}
\newcommand{\tj}{{\tilde{\mbox{\emph{\j}}}}}
\newcommand{\ts}{{\tilde{s}}}
\newcommand{\ti}{{\tilde{t}}}
\newcommand{\tups}{{\tilde{\upsilon}}}
\newcommand{\ttau}{{\tilde{\tau}}}
\newcommand{\tR}{{\tilde{R}}}
\newcommand{\bF}{{\mathbf{F}}}
\newcommand{\bv}{{\boldsymbol\upsilon}}

\newcommand{\bx}{{\mathbf{x}}}

\newcommand{\bn}{{\mathbf{n}}}
\newcommand{\btv}{{\tilde{\boldsymbol\upsilon}}}
\newcommand{\bta}{{\tilde{\boldsymbol a}}}
\newcommand{\btx}{{\tilde{\mathbf{x}}}}
\newcommand{\btj}{{\tilde{\mathbf{j}}}}
\newcommand{\btn}{{\tilde{\mathbf{n}}}}
\newcommand{\btR}{{\tilde{\mathbf{R}}}}

\newcommand{\EA}{\mathcal{E}_{\rm A}}
\newcommand{\EB}{\mathcal{E}_{\rm B}}

\newcommand{\longi}{{\mkern3mu\vphantom{\perp}\vrule depth 0pt\mkern3mu\vrule depth 0pt\mkern3mu}}

\begin{document}

\title{Radiation and a dynamical UV/IR connection in AdS/CFT}
\author{C\'esar A. Ag\'on}
\email[]{caagon87@brandeis.edu}
 \affiliation{Martin A. Fisher School of Physics, Brandeis University, Waltham, MA 02453, USA}
\author{Alberto G\"{u}ijosa}
\email[]{alberto@nucleares.unam.mx}
%\altaffiliation{Instituto de Ciencias Nucleares, Universidad Nacional
 % Aut\'onoma de M\'exico, Apartado Postal 70-543, M\'exico D.F. 04510, M\'exico}
\affiliation{Department of Physics, Princeton University, Princeton, NJ 08544, USA}
\thanks{On sabbatical leave from Instituto de Ciencias Nucleares, Universidad Nacional  Aut\'onoma de M\'exico}
\author{Juan F. Pedraza}
\email[]{jpedraza@physics.utexas.edu}
\affiliation{Theory Group, Department of Physics and Texas Cosmology Center, The University of Texas, Austin, TX 78712, USA}

\vspace{15pt}

\preprint{BRX-TH675}
\preprint{PUPT-2461}
\preprint{TCC-003-14}
\preprint{UTTG-03-14}

\begin{abstract}
{\bf Abstract:} We compute holographically the expectation value of the energy density sourced, in a strongly-coupled CFT, by a quark with large but finite mass (or equivalently, small but finite Compton radius) undergoing arbitrary motion. The resulting gluonic profile has two surprising features in the far region. First, besides the expected radiation, it contains a component that is attributable to the `intrinsic' or `near' field of the quark, and nevertheless falls off as the square of the distance. Second, even at distances much larger than the size of the quark, it differs from the profile set up by a pointlike quark. We explain how this second feature provides a useful case study for the UV/IR connection in a dynamical setting. We also examine some specific sample trajectories, including uniform circular motion and harmonic oscillation, where features such as the extent of the region with negative energy are found to vary with the quark mass.
 \end{abstract}

\maketitle

\tableofcontents

\section{Introduction and Conclusions}\label{introsec}

\subsection{Motivation}\label{motivationsubsec}
One of the basic aspects of interest in any field theory is the way in which localized sources can give rise to radiation, or more generally, to propagating disturbances in the fields. If the theory is strongly-coupled, this is a question about which traditional methods give little information, so it is natural that the efficient tools of holography \cite{malda,gkpw} have been brought to bear on it on many occasions, starting with the early works \cite{bklt,dkk,cg}. Our interest will lie here on radiation by a pointlike or nearly-pointlike source in the vacuum of a $d$-dimensional conformal field theory (CFT, with large central charge and a large gap in the spectrum of conformal dimensions), which is known to be dual to a string on $d+1$-dimensional anti-de Sitter (AdS) times a compact space. For concreteness, we will phrase our discussion in terms of the correspondence between $\cN=4$ $SU(N)$ super-Yang-Mills (SYM) theory on $3+1$ Minkowski space and Type IIB string on (a Poincar\'e wedge of) AdS$_5\times\bS^5$, where the source is a heavy quark (a fundamental hypermultiplet) and the CFT fields will be generically called gluonic. The simplest observables that can serve to map out the profile of these fields are one-point functions of local operators such as the energy-momentum tensor $T_{\mu\nu}$ or the Lagrangian density $\tr{F^2}+\ldots$

Recent works in this context have found a number of interesting properties, including the fact that, in spite of the strong coupling, field disturbances do not display temporal (or equivalently, radial) broadening \cite{iancu1,HIMT,trfsq}, and can be beamed if the motion is relativistic \cite{liusynchrotron,veronika,beaming}.
Our exploration here will build particularly on the results of Hatta, Iancu, Mueller and Triantafyllopoulos \cite{HIMT}, who (following in turn \cite{liusynchrotron}) derived the expectation value of the energy density, $\cE(x)\equiv\expec{T_{00}(x)}$, in the presence of an infinitely massive quark undergoing \emph{arbitrary} motion. The key ingredient that made this calculation possible is knowledge of the full, nonlinearized string embedding dual to such a quark, provided in earlier work by Mikhailov \cite{mikhailov}. The calculation proceeds by first determining the bulk gravitational field generated by this string and then considering its behavior near the AdS boundary, to extract $\cE(x)$.

 There exists an alternative approach to determine the energy of the gluonic fields in this setup. By directly computing the energy of the string described by his solution, Mikhailov demonstrated that the quark's rate of radiation coincides with the Lienard formula familiar from classical electromagnetism. His results, in combination with those of \cite{dragtime}, showed how to split the total energy of the string into two contributions that can be respectively recognized as the intrinsic energy of the quark at that instant and the total energy that has been carried away as radiation at all previous times. It is natural to try to make contact between the outcome of this worldsheet analysis and the result for $\cE(x)$ afforded by the bulk approach described in the previous paragraph. The latter provides detailed information about the spacetime distribution of the energy, but on the other hand, \emph{a priori} it does not distinguish between intrinsic and radiated energy. Resorting to the usual definition of radiation as the component of the field that can transport energy far away from the source, the authors of \cite{HIMT} examined $\cE(x)$ at large distances, and found a mismatch with the worldsheet (i.e., Lienard) result. Moreover, the unexpected additional terms in the far zone energy density were seen to have various properties that would be rather peculiar if they were to be interpreted as encoding radiation. In fact, their structure was found to be compatible instead with a piece of the \emph{intrinsic} quark energy obtained via the worldsheet approach in \cite{dragtime,lorentzdirac,damping}.

Importantly, the analysis of \cite{dragtime,lorentzdirac,damping} was carried out for a quark with large but finite mass, yielding a rate of radiation that differs from Lienard and an intrinsic quark energy that is no longer just $\gamma m$. So, to make a valid comparison, the results of \cite{HIMT} must be upgraded to finite mass as well. This is the main task that we carry out in the present paper. Beyond the immediate motivation of trying to resolve the puzzle just described, considering the case with $m<\infty$ is interesting in itself, for a number of reasons.
First, it is of course more realistic.
Second, a finitely-massive quark is automatically dressed with a `gluon cloud' of finite size \cite{martinfsq,damping}, and it is remarkable that the AdS/CFT correspondence grants us easy access to the dynamics of this extended object, manifested, e.g., in its nontrivial dispersion relation.
 Third, there are other physical effects, such as radiation damping, that are only visible at finite quark mass, and again are only rendered manageable at strong coupling through the power of AdS/CFT  \cite{lorentzdirac,damping}. Fourth, on the gravity side, the finite size of the quark translates into the fact that the dual string does not extend all the way to the AdS boundary. As we will explain below, this provides a useful dynamical probe of the mapping between bulk and boundary regions, and more specifically, of the UV/IR connection \cite{uvir} that is a cornerstone of the AdS/CFT correspondence and a practical tool in many holographic arguments.

\subsection{Summary of results}\label{summarysubsec}

 The organization of the paper is as follows. In section \ref{mikhsec} we review the worldsheet approach: Mikhailov's solution \cite{mikhailov}, its extension to finite quark mass \cite{dragtime}, and the ensuing equation of motion (\ref{eom}) for the dressed quark, from which covariant expressions for its dispersion relation (\ref{pq}) and radiation rate (\ref{radiationrate}) can be read off \cite{lorentzdirac,damping}. In section \ref{bulkstresssec} we assemble the ingredients needed for the bulk approach: the bulk stress tensor sourced by a heavy quark following an arbitrary trajectory \cite{HIMT}, and the general formula for the CFT energy density \cite{liusynchrotron,HIMT} that follows from the corresponding backreacted bulk geometry.

 The main calculation of this paper is then carried out in Section \ref{densitysec}, arriving at (\ref{e12}), where, in preparation for the subsequent analysis, the energy density is presented as a sum of two contributions, $\cE=\cE^{(1)}+\cE^{(2)}$, that include all terms respectively without and with overall time derivatives. The profile again shows no temporal/radial broadening: its value at a given observation point is seen to depend only on the behavior of the quark at a single retarded time determined by (\ref{tauret}). This is the same time that was previously found to be relevant for disturbances in $\expec{\tr{F^2}+\ldots}$ \cite{trfsq}, and it implies a subluminal propagation speed.

 In Section \ref{totalsec} we make some comments about the process of integrating $\cE$ over space to determine the total energy at a fixed observation time, and ultimately run into expressions that are too unwieldy. The principal lesson we gain from this exercise is that the natural conjecture that $\cE^{(1)}$ and $\cE^{(2)}$ might respectively match upon integration with the radiated and intrinsic energy deduced from the worldsheet approach is \emph{not} correct.

  We then extract concrete information from our general formulas by considering some specific sample trajectories: the static and uniform velocity cases in Sections \ref{staticsubsec} an \ref{constantsubsec}, where all energy is of course intrinsic, as well as uniform circular motion in Section \ref{circularsubsec} and harmonic oscillation in \ref{harmonicsubsec}, where radiation is present. In the accelerated cases various features of the profile, including the strength and extent of the regions where the energy is negative, are seen to vary as the quark mass is decreased. We also point out in Section \ref{forcedrestsubsec} that, contrary to what was believed in \cite{damping}, the solution to the equation of motion (\ref{eom}) where the quark is forced but remains at rest is ultimately unphysical. More generally, there is just one possible external forcing that gives rise to any given quark motion.

  In Section \ref{farsec} we go back to the case of arbitrary motion and examine the gluonic profile at large distances $R$ from the quark, where vast simplifications occur. As expected, the dominant falloff is $1/R^2$ and arises from terms that involve the acceleration. Following \cite{HIMT}, we work out the total power flowing through the sphere at infinity for a fixed emission time at the source. For any value of the quark mass, the contribution from $\cE^{(1)}$, presented in (\ref{P1rad}), is found to be in complete agreement with the rate of radiation (\ref{radiationrate}) inferred from the worldsheet approach. This unequivocally implies that if the contribution from $\cE^{(2)}$ is nonvanishing, it must necessarily be due to the intrinsic or `near' component of the field. The first few terms in an expansion in (inverse) powers of the quark size (mass) are shown in (\ref{P2rad}). The lowest-order term agrees with, and thereby validates, the result of \cite{HIMT}. The clearest sign that (\ref{P2rad}) indeed arises from the component of the field that is still adhered to the quark is the fact that, unlike (\ref{P1rad}), it can be negative, indicating that by appropriately tugging on the quark we can pull this energy \emph{inward} from infinity!

  Our results thus confirm the suspicion of \cite{HIMT} that in this system radiation is \emph{not} synonymous with the $1/R^2$ component of the gluonic field, because there is a tail of the `near' field that extends all the way out to the far zone. This surprising fact is related to a recent finding of Lewkowycz and Maldacena \cite{lewkomalda}. For the case of an infinitely heavy quark undergoing uniform acceleration, these authors noticed a discrepancy between the coefficient in the exact one-point function of the energy-momentum tensor and the so-called Bremsstrahlung function \cite{correa,fiol} that determines, among other things, the quark's rate of radiation. They attributed the discrepancy to the need to properly separate the intrinsic and radiative components of the energy, a task which,
for uniform acceleration, is known to be difficult even in classical electromagnetism.
 Lewkowycz and Maldacena then went on to show that an invariant subtraction procedure, based on the fact that the quark also sources a dimension 2 scalar operator, successfully removes the discrepancy. In other words, at least for this type of motion, the portion that is subtracted is precisely the intrinsic contribution. As we discuss at the end of Section \ref{farsec}, our results shed light on this issue, and show that the discrepancy encountered in \cite{lewkomalda} arises not from the peculiar nature of uniform acceleration, but from the presence of a $1/R^2$ tail of the intrinsic energy for arbitrary (nonstationary) trajectories. In the opposite direction, the results of \cite{lewkomalda} teach us that this tail is not an effect that is exclusive to the strong-coupling regime, but rather originates from the fact that our quark excites a profile in the conformally coupled scalar fields of the gauge theory.

  We should stress that, while in Section \ref{farsec} we find convincing evidence that $\cE^{(1)}$ and $\cE^{(2)}$ respectively encode the radiative and intrinsic components of the gluonic field when examined at large distances for fixed emission time, from Section \ref{totalsec} we know that this identification does \emph{not} hold when these 2 contributions of (\ref{e12}) are examined at arbitrary spacetime locations. If we were able to perform the split between these two components of the gluonic profile at each spacetime point, then we could quantify radiation locally, without going far away from the source. Of course, while such a local split can be performed in classical electromagnetism \cite{teitelboim}, it might well prove impossible to achieve in the highly nonlinear setting of a strongly-coupled non-Abelian gauge theory. But the procedure of \cite{lewkomalda} seems to be efficiently achieving precisely this, at least for the case of an infinitely massive quark undergoing uniform acceleration. A very interesting question then, which we leave for future work, is whether it succeeds in providing us with this separation even for arbitrary motion and  finite quark mass.\footnote{Notice in particular that the scalar fields do in fact make a contribution to the radiative component \cite{liusynchrotron,HIMT}, which the prescription of \cite{lewkomalda} must then \emph{not} subtract.}

At the end of the paper, we turn our attention to another surprising feature of our result for the energy profile. Let $z_m>0$ denote the finite size that the quark acquires on account of its having a finite mass (the precise relation is (\ref{zm})). Our intuition would tell us that, if we analyze the gluonic fields only at distances much greater than this size, $R\gg z_m$, then we should not be able to tell that the quark is not pointlike. In the gravity description, the scale $z_m$ marks the radial location in AdS at which the string terminates. In the standard coordinates where the  metric takes the form (\ref{metric}), the AdS boundary is at $z=0$, and the string lives entirely in the region $z\ge z_m$. Through the UV/IR connection \cite{uvir}, we know that the near-boundary region of the bulk that the string does not reach corresponds to the UV of the gauge theory, which is consistent with the expectation that at long distances we should not be able to notice that $z_m>0$.

 In Section \ref{uvirsec}, we use our results and those of \cite{trfsq} to put this expectation to the test. Before describing our findings, we should point out that exactly this same test is relevant in the following alternative scenario. A quark that is truly pointlike (and therefore infinitely massive) is dual to a string that does reach the AdS boundary. But suppose that, for some reason, we only have information on the profile of this string in some region $z\ge z_m$ that does not extend all the way to the boundary. A concrete example is the interesting proposal put forth by Hubeny \cite{veronika} to provide a gravity-side explanation of the beaming of gluonic radiation, by approximating the string as a collection of point sources that set up gravitational shock waves. This approximation is valid as long as certain conditions are met \cite{veronika,beaming}, which in general happens only beyond some radial distance $z_m$ away from the AdS boundary (where the value of $z_m$ depends on the behavior of the quark).\footnote{An alternative beaming mechanism was found in \cite{beaming}, which is valid independently of any approximation, and could well operate jointly with the shock wave construction of \cite{veronika}.}
 Again, by the standard UV/IR reasoning, we would expect the missing near-boundary information not to matter if we retreat to distances $R\gg z_m$ away from the quark.

 In either scenario, our test bears upon locality properties of the AdS/CFT correspondence that have been explored by different means in a number of other works. In particular, the recent papers \cite{bousso,mark,mukund,vijay,wall} analyzed issues related to the following question: given full access to data only on a finite region of the underlying CFT spacetime, how deep into the bulk is it possible to reconstruct the dual background? The issue that
 we address is to some extent the complement of this question, because we are interested in identifying the CFT region that we lose control over when we \emph{do not} have access to bulk data within a given radial distance $z_m$ from the AdS boundary.\footnote{Of course, the details are different, because our analysis relates specifically to data flow along a string associated with a moving quark, whereas previous studies explored the flow of information through the bulk of the (asymptotically) AdS geometry.}

  When the quark is accelerating, the answer we find in Section \ref{uvirsec} is not in line with the naive UV/IR expectation: the mapping from bulk to boundary turns out to be sufficiently nonlocal that, even at distances
  $R\gg z_m$ away from the quark, the missing or omitted portion of the string turns out \emph{not} to be negligible, unless the state of motion of the quark does not change appreciably over a time interval of order $z_m$, as expressed in (\ref{nonviolent3}). The logic of this restriction is explained in the paragraphs that follow it, and in retrospect, seems rather natural. The key point is that, in our time-dependent system, relinquishing UV information on the source not only limits the spatial region of the gluonic field that we can access---it also places a limit on our \emph{temporal resolution}. It is this dynamical aspect of the UV/IR connection that prevents us from being sensitive to motion of the quark that is too abrupt.

\section{String Embedding and Quark Equation of Motion} \label{mikhsec}
The geometry dual to the (symmetry-preserving)
vacuum of $\cN=4$ SYM
on $(3+1)$-dimensional Minkowski space is AdS$_5$ in Poincar\'e slicing,\footnote{Cross a sphere $\mathbf{S}^5$, which will not play any role in our discussion. Replacing it by a different compact space $\mathbf{X}^5$ corresponds to replacing $\cN=4$ SYM by a different $(3+1)$-dimensional CFT.
Our results are valid for any of such cases as long as
they are expressed in terms of the appropriate CFT coupling, which will differ from (\ref{lambda}) because this relation depends on the geometry of $\mathbf{X}^5$. The extension to CFTs in other dimensions is also straightforward.}
\begin{equation}\label{metric}
 \rmd s^2\,\equiv\,G_{mn}\,\rmd x^m \rmd x^n\,= \,\frac{L^2}{z^2} \,\left( \eta_{\mu\nu}\rmd x^\mu\rmd x^\nu + \rmd z^2 \right).
\end{equation}The radius of curvature $L$ is related to the SYM 't Hooft coupling $\lambda\equiv g_{YM}^2 N_c$ through
\begin{equation}\label{lambda}
 \lambda=\frac{L^4}{l_s^4}~,\nonumber
\end{equation}
where $l_s$ denotes the string length.

A quark with mass $m$ is dual to a string that on one side reaches the Poincar\'e horizon $z\to\infty$ and on the other
ends on a stack of flavor D7-branes at $z=z_m$, with
\begin{equation}\label{zm}
z_m=\frac{\sqrt{\lambda}}{2\pi m}~.
\end{equation}
The string dynamics is prescribed as usual by the Nambu-Goto action
\begin{equation}\label{nambugoto}
S_{\text{NG}}=-T_0\int
d^2\sigma\,\sqrt{-\det{g_{ab}}}~,
\end{equation}
where $T_0=1/2\pi l_s^2$ is the string tension and $g_{ab}\equiv\partial_a X^m\partial_b X^n G_{mn}(x)$  denotes
the induced metric on the worldsheet.

Choosing $(\tau,z)$ as the two coordinates parametrizing
the worldsheet, Mikhailov's solution for the string embedding dual to an infinitely massive quark, $z_m=0$, can be written as \cite{mikhailov}
\begin{equation}
X^m = (X^\mu(\tau,z),z),
\end{equation}where
\be\label{MikSol}
X^\mu(\tau,z)=x^\mu(\tau)+z\upsilon^\mu(\tau).
\end{equation}Here $x^\mu$ represents the worldline of the quark, or equivalently, of the string endpoint at $z=0$, parameterized by its proper time $\tau$, and $\upsilon^\mu\equiv dx^\mu/d\tau$ is its four-velocity, with $\eta_{\mu\nu}\upsilon^\mu \upsilon^\nu=-1$.
For this solution, the induced metric on the worldsheet turns out to be
\begin{equation}\label{wsmetric} g_{\tau\tau}={L^2\over
z^2}(z^2a^2-1),\qquad g_{zz}=0,\qquad g_{z\tau}=-{L^2\over z^2},
\end{equation}
where $a^2$ is the square of the four-acceleration $a^\mu\equiv d^2 x^\mu/d^2\tau$. This worldsheet metric implies that the lines with constant $\tau$ are null, a fact that plays an
important role in Mikhailov's construction. In particular, the solution (\ref{MikSol}) is \emph{retarded}, in the sense
that the behavior at any $t=X^{0}(\tau,z)$ of the string segment located at a radial position $z$ is
completely determined by the behavior of the string endpoint at an earlier time $t_r(t,z)$
obtained by projecting back towards the boundary along the null line at fixed $\tau$. In gauge theory language, this
condition corresponds to a purely outgoing boundary condition for the waves in the gluonic field at spatial infinity.

For the case of a quark with finite mass, $z_m>0$, the string embedding  can be regarded as the $z\ge z_m$ portion of the solution
(\ref{MikSol}),
which, as mentioned before, is parametrized by data at the AdS boundary $z=0$. We will use tildes to label
these auxiliary data, and distinguish them from the actual physical quantities
(velocity, proper time, etc.) associated with the endpoint/quark at $z=z_m$, which will be denoted
without tildes.  In this notation, the solution reads
\begin{equation}\label{mikhsoltilde}
X^{\mu}(\ttau,z)=\tx^{\mu}(\ttau)+z\tilde{\upsilon}^{\mu}(\ttau)~.
\end{equation}
To get the quark/endpoint to follow a given trajectory, we must exert on it
an external force $\cF^{\mu}=(\gamma\vec{F}\cdot\vec{v},\gamma\vec{F})$. In the gravity description this can be accomplished by turning on the electromagnetic field $A_{\mu}$ on the flavor D-branes where the string ends. It amounts to adding to the Nambu-Goto action the usual minimal coupling
\begin{equation}\label{couplingtoA}
S_{\text{F}}=\int
\dif\tau\,A_{\mu}(X(\tau,z_m))\partial_{\tau}X^{\mu}(\tau,z_m)~,
\end{equation}where $\tau$ is now the true proper time for the string endpoint $X^{\mu}(\tau,z_m)=x^\mu(\tau)$.
Variation of the string action $S_{\text{NG}}+S_{\text{F}}$ implies the
standard Nambu-Goto equation of motion for all interior points of the string, as well as the
boundary condition
\begin{equation}\label{stringbc}
\Pi^{z}_{\mu}(\tau)|_{z=z_m}=\cF_{\mu}(\tau)\quad\forall~\tau~,
\end{equation}where
\begin{equation}\label{pizmu}
\Pi^{z}_{\mu}\equiv \frac{\partial\mathcal{L}_{\text{NG}}}{\partial(\partial_z
X^{\mu})}
\end{equation}is the worldsheet Noether current associated with spacetime momentum, and $\cF_{\mu}=F_{\mu\nu}\upsilon^{\nu}$ is the Lorentz force.

In the present paper we will find it convenient to carry out the computations in terms of the auxiliary variables $\tilde{x}^\mu$, $\tilde{\upsilon}^\mu$, etc., and only at the end restate our results in terms of the real physical data $x^\mu$, $\upsilon^\mu$, etc. The connection between these two sets of variables is given by \cite{damping}
\begin{equation}\label{tau2}
 \dif\tilde{\tau}=\frac{\dif\tau}{\sqrt{1-z^4_m \bar{\cF}^2}}~,
\end{equation}
\begin{equation}\label{xtilde}
\tx^{\mu}=x^{\mu}-z_m\frac{\upsilon^\mu-z^2_m\bar{\cF}^{\mu}}
{\sqrt{1-z^4_m \bar{\cF}^2} }~,
\end{equation}
\begin{equation}\label{vtilde}
\tilde{\upsilon}^{\mu}=\frac{\upsilon^\mu-z^2_m\bar{\cF}^{\mu}}{\sqrt{1-z^4_m \bar{\cF}^2} }~,
\end{equation}
\begin{equation}\label{atilde}
\ta^{\mu}=z_m\frac{\bar{\cF}^{\mu}-z^2_m \bar{\cF}^2\upsilon^\mu}{\sqrt{1-z^4_m \bar{\cF}^2}}~,
\end{equation}
etc., where $\bar{\cF}_{\mu}\equiv (2\pi/\sqrt{\lambda})\cF_{\mu}$.
These relations imply in particular that the string embedding (\ref{mikhsoltilde}) can be written purely in terms of physical variables as
\begin{equation}\label{mikhsolzm}
X^{\mu}(\tau,z)=x^{\mu}(\tau)+ \frac{(z-z_m)(\upsilon^{\mu}-z^2_m\bar{\cF}^{\mu})}{\sqrt{1-z^4_m \bar{\cF}^2}}~.
\end{equation}

The equation of motion for the quark itself comes from the boundary condition (\ref{stringbc}) and can be written in terms of the physical data as
\cite{lorentzdirac,damping}
\begin{equation}\label{eom}
{d\over
d\tau}\left(\frac{m\upsilon^{\mu}-{\sqrt{\lambda}\over 2\pi m}
\cF^{\mu}}{\sqrt{1-{\lambda\over 4\pi^2 m^4}\cF^2}}\right)=\frac{\cF^{\mu}-{\sqrt{\lambda}\over 2\pi
m^2} \cF^2 \upsilon^{\mu}}{1-{\lambda\over 4\pi^2 m^4}\cF^2}~.
\end{equation}Remarkably, in terms of the tilde variables this is just a restatement of the purely kinematic equation $\dif \tilde{\upsilon}^\mu/\dif \tilde{\tau}=\tilde{a}^\mu$ \cite{dampingtemp}.

For the interpretation of our results, it is important to notice that (\ref{eom}) can be rewritten as a statement of the rate of change of the total four-momentum of the system \cite{lorentzdirac,damping},
 \begin{equation}\label{eomsplit}
 {d p^{\mu}\over d\tau}\equiv {d p_q^{\mu}\over d\tau}+{d p^{\mu}_{\mbox{\scriptsize rad}}\over
 d\tau}=\cF^{\mu},
 \end{equation}
 where
 \begin{equation}\label{pq}
 p_q^{\mu}=\frac{m\upsilon^{\mu}-{\sqrt{\lambda}\over 2\pi m}
 \cF^{\mu}}{\sqrt{1-{\lambda\over 4\pi^2 m^4}\cF^2}}
 \end{equation}
 is recognized as the intrinsic four-momentum of the quark (including all near-field contributions), and
\begin{equation}\label{radiationrate}
{d p^{\mu}_{\mbox{\scriptsize rad}}\over
d\tau}={\sqrt{\lambda}\, \cF^2 \over 2\pi m^2}\left(\frac{\upsilon^{\mu}-{\sqrt{\lambda}\over
2\pi m^2} \cF^{\mu} }{1-{\lambda\over 4\pi^2 m^4}\cF^2}\right)
\end{equation}
represents the rate at which four-momentum is carried away from the quark by gluonic radiation. It is easy to check that $p_q^2=-m^2$, which makes it clear that the splitting given by (\ref{eomsplit}) is correctly Lorentz covariant.

The non-standard form of (\ref{pq}) and (\ref{radiationrate}) reflects the fact that, for $m<\infty$, the quark is no longer pointlike, and develops a `gluonic cloud' \cite{martinfsq,damping} with characteristic
 size $\sqrt{\lambda}/2\pi m$. In other words, the radial AdS scale
 $z_m$ given by (\ref{zm}) has a direct physical interpretation in the gauge theory: it plays the role of the Compton radius of the quark.\footnote{Notice that $z_m$ is a factor of $\sqrt{\lambda}$ larger than the naive Compton radius $1/m$ that would be relevant at weak coupling. In our strongly-coupled setting, the size of the virtual cloud for the quark is set by the mass scale \cite{martinmeson,strassler} for the \emph{deeply bound} mesons , $m_{\mbox{\scriptsize mes}}=m/\sqrt{\lambda}$.} For $m\to\infty$, we recover the expected pointlike dispersion relation $p_q^{\mu}=m\upsilon^{\mu}$ and Lienard rate of radiation
 ${d p^{\mu}_{\mbox{\scriptsize rad}}/d\tau}=\sqrt{\lambda}a^2\upsilon^{\mu}/2\pi$, which were first deduced respectively in \cite{dragtime} and \cite{mikhailov}, by examining the total four-momentum of the string at a fixed  time $t$. For finite quark mass, the split (\ref{eomsplit}) into intrinsic and radiated four-momentum was first obtained in \cite{dragtime}, in the case of motion purely along one-dimension.  A prominent feature of the dispersion relation
  (\ref{pq}) and radiation rate (\ref{radiationrate}) is the appearance of a divergence when
  $\cF^2=\cF^2_{\mbox{\scriptsize crit}}$, where
  \begin{equation}\label{Fcrit}
  \cF^2_{\mbox{\scriptsize crit}}={ 4\pi^2 m^4\over\lambda}
  \end{equation}
  ($\bar{\cF}^2_{\mbox{\scriptsize crit}}=1/z_m^4$) is the critical value at which the force can nucleate quark-antiquark pairs,
  or, in gravity language, create open strings \cite{ctqhat}.

Conceptually, the factors of the external force that appear in the dispersion relation (\ref{pq}) and radiation rate (\ref{radiationrate}) should be understood as just a convenient way to summarize the dependence of these quantities on the variables describing the motion of the quark. Indeed, as long as the force is subcritical, the equation of motion (\ref{eom}) can be rewritten in the form of a derivative expansion \cite{damping,lessons}
 \begin{equation}\label{force}
 \cF^{\mu}=ma^{\mu}+\frac{\sqrt{\lambda}}{2\pi}(a^2 \ups^{\mu}-j^{\mu})
 +\frac{\lambda}{4\pi^2 m}(s^{\mu}-3a\cdot j \ups^{\mu}-\frac{3}{2}a^2 a^{\mu})
 +\ldots
 \end{equation}
involving the quark's velocity $\ups^{\mu}$, acceleration $a^{\mu}$, jerk $j^{\mu}\equiv d^3 x^{\mu}/d\tau^3$, snap $s^{\mu}\equiv d^4 x^{\mu}/d\tau^4$, and all higher derivatives.\footnote{See \cite{tr} for an interesting discussion of the jerk, snap, etc. in the relativistic context.} Keeping only the terms up to order $m^0$, this relation matches the classic Lorentz-Dirac equation (constructed to incorporate the effects of radiation damping), so (\ref{eom}) can be recognized as a physically sensible nonlinear generalization thereof \cite{lorentzdirac}.\footnote{Damping of the quark is a first effect of the emission of radiation, which is already visible at the level of the classical description of the string. Quantum mechanically, the emitted radiation additionally induces stochastic fluctuations in the quark trajectory, which have been studied in \cite{Xiao:2008nr,hirayama,Caceres:2010rm,Fiol:2013iaa}.} Some of its physical implications were explored in \cite{damping,kiritsis}.

\section{Bulk Stress Tensor and Backreaction} \label{bulkstresssec}

 The contribution of the string to the stress tensor in the AdS bulk is
 \begin{eqnarray}\label{Tmn}
\mathrm{T}^{mn}(x)\, &=&\,\frac{2}{\sqrt{-G}}\frac{\delta S_{\text{NG}}}{\delta G_{mn}(x)}\nn\, &=&\,
 -\frac{T_0}{\sqrt{-G}}\int \rmd^2\sigma~\sqrt{-g}\,g^{ab}\,
 \partial_a X^m \,\partial_b X^n\,
 \delta^{(5)}(x-X(\tau,\sigma))\\
 \, &=&\, \int \rmd \ttau ~\mathrm{t}^{mn}\, \delta^{(4)}(x^\mu-X^\mu(\ttau,z))\nonumber,
 \end{eqnarray}
 where
 \begin{equation}
 \mathrm{t}^{mn} = \frac{T_0}{\sqrt{-G}\sqrt{-g}}\,
 \left[
 g_{\tau \tau} {X^m}' {X^n}' -
 g_{\tau z} \left(\dot{X}^m {X^n}' + {X^m}' \dot{X}^n\right)
 \right].
 \end{equation}In particular, we find that
 \begin{eqnarray}
 \mathrm{t}^{\mu\nu} &=& T_0 \frac{z^5}{L^5}\left[\left(z^2\ta^2+1\right)\tups^\mu\tups^\nu
+z\left(\ta^\mu\tups^\nu+\tups^\mu \ta^\nu\right)\right],
 \nonumber\\
\mathrm{t}^{\mu z} &=& T_0 \frac{z^5}{L^5}\left(z^2\ta^2\tups^\mu+z \ta^\mu\right),
 \\
\mathrm{t}^{zz} &=& T_0 \frac{z^5}{L^5}\left(z^2\ta^2-1\right).
\nonumber
\end{eqnarray}
We now want to consider the backreaction of the string on the geometry, so that we can make use of the GKPW recipe for
correlation functions \cite{gkpw} to compute the expectation value of the energy density $\mathcal{E}\equiv\left\langle T_{00}\right\rangle$ on the boundary. This is essentially the convolution of $\mathrm{T}^{mn}$ with
the graviton bulk-to-boundary propagator. The final result can be summarized as \cite{liusynchrotron,HIMT}
 \begin{eqnarray} \mathcal{E}(x^\mu) \!&=&\! \EA(x^\mu)+\EB(x^\mu),
 \end{eqnarray}
 where the two contributions are
 \begin{eqnarray}\label{EA}
 \EA \!&=&\! \frac{2 L^3}{\pi}\!
 \int\! \frac{\dif^4x' \,\dif z}{z^2}\, \Theta(t-t') \delta''(\mathcal{W})\!
 \left[z (2 \mathrm{T}_{tt} - \mathrm{T}_{zz}) - (t-t') \mathrm{T}_{tz} + (x - x')^i \mathrm{T}_{iz}
 \right],
\\ \label{EB}
 \EB\!&=&\! \frac{2 L^3}{3 \pi}\!
 \int\! \frac{\dif^4x' \,\dif z}{z}\, \Theta(t-t') \delta'''(\mathcal{W})\!
 \left[(\mathbf{x} - \mathbf{x}')^2 (2 \mathrm{T}_{tt} -2 \mathrm{T}_{zz} + \mathrm{T}_{ii}) -
 3 (x - x')^i (x - x')^j \mathrm{T}_{ij} \right].
 \nonumber
 \end{eqnarray}
 Here, the argument of $\mathrm{T}_{mn}$ is $(t',\mathbf{x}',z)$ and the quantity
 \begin{equation}\label{W}
 \mathcal{W} \equiv (x-x')^\mu(x-x')_\mu+z^2=-(t-t')^2 + (\mathbf{x} - \mathbf{x}')^2 +z^2\,
 \end{equation}is proportional to the 5D invariant distance between the source point in
the bulk and the measurement point on the boundary.

\section{Energy Density sourced by the Heavy Quark}\label{densitysec}

We will follow \cite{HIMT} closely. Let us first focus on the $\EA$ part of the energy density. The integration over $\dif^4x'$ is trivially done using the $\delta$-function of the string stress tensor (\ref{Tmn}), and after some algebra we obtain
 \begin{equation}\label{EAtemp}
 \EA = \frac{\sqrt{\lambda}}{\pi^2}
 \int \dif \ttau\,\dif z\, \delta''(\mathcal{W}_q + 2 z \Xi)
 \left[A_0(\ttau) + z A_1(\ttau)\right],
 \end{equation}with the definitions
 \begin{equation}\label{Wq}
 \mathcal{W}_q \equiv (x-\tx(\ttau))^\mu(x-\tx(\ttau))_\mu~,
 \qquad
 \Xi \equiv
 -(x-\tx(\ttau))^\mu\tups_\mu(\ttau)=
 \,\frac{1}{2}\,\frac{\dif \mathcal{W}_q}{\dif \ttau}~,
 \end{equation}
 and where the coefficients inside the square bracket are
 \begin{eqnarray}\label{As}
 A_0(\ttau) &=& 1-2\tups^0\tups_0+\left[-\left(t-\ti\right)\ta_0+\left(x-\tx\right)^i\ta_i\right],
 \\
 A_1(\ttau) &=& -2\ta^0\tups_0+\ta^2\left[-\left(t-\ti\right)\tups_0+\left(x-\tx\right)^i\tups_i\right].
 \nonumber
 \end{eqnarray}
 The argument of the $\delta$-function has become linear in $z$ and thus we can start either by performing the corresponding integral, or by changing variables to $(\ttau,z)\rightarrow(\ttau,\mathcal{W})$. The second path is neater for us, though one can show that the results are equal, as they should be. We then proceed by making the substitutions
\begin{equation}
z=\frac{\mathcal{W}-\mathcal{W}_q}{2\Xi}
\end{equation}
and
\begin{equation}
\dif z=\frac{1}{2\Xi}\dif\mathcal{W}~.
\end{equation}
The integral over $z$ runs from $z=z_m$ up to $z\rightarrow\infty$. To make this explicit we introduce a step function $\Theta(z-z_m)$ in the integrand, which in terms of the new variables becomes $\Theta(\mathcal{W}-\mathcal{W}_q-2z_m\Xi)$. We then integrate over the full domain of $\mathcal{W}$,
\begin{equation}\label{EAtemp2}
 \EA = \frac{\sqrt{\lambda}}{\pi^2}
 \int \dif \ttau\,\dif \mathcal{W}\, \delta''(\mathcal{W})\Theta(\mathcal{W}-\mathcal{W}_q-2z_m\Xi)
 \left[\frac{A_0}{2\Xi} + \frac{A_1}{4\Xi^2}(\mathcal{W}-\mathcal{W}_q)\right]~.
\end{equation}
In order to integrate this by parts it is convenient to rewrite the argument as a function of the combination $f(\mathcal{W}-\mathcal{W}_m)$, where $\mathcal{W}_m\equiv\mathcal{W}_q+2z_m\Xi$. After some algebra, expression (\ref{EAtemp2}) can be massaged into the form
\begin{eqnarray}
\EA &=& \frac{\sqrt{\lambda}}{\pi^2}
 \int \dif \ttau\,\dif \mathcal{W}\, \delta''(\mathcal{W})\Theta(\mathcal{W}-\mathcal{W}_m)\left[
 \frac{A_0 +z_m A_1}{2\Xi}+\frac{A_1}{4\Xi^2}(\mathcal{W}-\mathcal{W}_m)\right],\\
 &=&
 \frac{\sqrt{\lambda}}{\pi^2}
 \int \dif \ttau\,\dif \mathcal{W}\, \delta(\mathcal{W})\left\{\frac{A_0 +z_m A_1}{2\Xi}\frac{\partial^2\,}{\partial \mathcal{W}^2}\Theta(\mathcal{W}-\mathcal{W}_m)+\frac{A_1}{4\Xi^2}\frac{\partial^2\,}{\partial \mathcal{W}^2}\left[(\mathcal{W}-\mathcal{W}_m)\Theta(\mathcal{W}-\mathcal{W}_m)\right]\right\},\nonumber
\end{eqnarray}
where the partial derivatives are taken holding $\ttau$ constant. One can
 check that the integration by parts generates no boundary terms.
Now, for any function of the form $f(\mathcal{W}-\mathcal{W}_m)$,
$$
\frac{\partial f}{\partial \mathcal{W}}=-\frac{\partial f}{\partial \mathcal{W}_m}\,,
$$
so the derivatives can now be taken with respect to $\mathcal{W}_m$ and pulled outside the $\mathcal{W}$-integration.
Using the $\delta$-function we then get
\begin{eqnarray} \label{inttao}
\EA &=& \frac{\sqrt{\lambda}}{\pi^2}
 \int \dif \ttau\, \left\{\frac{A_0 +z_m A_1}{2\Xi}\frac{\partial^2\,}{\partial \mathcal{W}_m^2}\Theta(-\mathcal{W}_m)-\frac{A_1}{4\Xi^2}\frac{\partial^2\,}{\partial \mathcal{W}_m^2}\left[\mathcal{W}_m\Theta(-\mathcal{W}_m)\right]\right\},
 \\
 &=& \frac{\sqrt{\lambda}}{\pi^2}
 \int \dif \ttau\, \left[-\frac{A_0 +z_m A_1}{2\Xi}\frac{\partial}{\partial \mathcal{W}_m}\delta(\mathcal{W}_m)+\frac{A_1}{4\Xi^2}\delta(\mathcal{W}_m)+\frac{A_1}{4\Xi^2}\frac{\partial}{\partial \mathcal{W}_m}\left[\mathcal{W}_m\delta(\mathcal{W}_m)\right]\right].
 \nonumber
\end{eqnarray}
At this point we can make use of general properties of $\delta$-functions to write
\begin{equation}
\delta(\mathcal{W}_m)=\frac{\delta(\ttau-\ttau_r)}{2|\Xi_m|},
\end{equation}where we have defined
\begin{equation}\label{Xim}
\Xi_m\equiv\frac{1}{2}\,\frac{\dif\mathcal{W}_m}{\dif \ttau}=\Xi-z_m\left[1+(x-\tx)^\mu \ta_\mu\right]\,,
\end{equation}and where $\ttau_r$ is the unique
 solution to the equation (to be discussed later)
\begin{equation}\label{tauretnew}
\mathcal{W}_m(\ttau_r)\equiv\mathcal{W}_q(\ttau_r)+2z_m\Xi(\ttau_r)=0.
\end{equation}We also need the fact that
\begin{equation}
\frac{\partial}{\partial \mathcal{W}_m}\delta(\mathcal{W}_m)=\frac{1}{2\Xi_m}\frac{\dif\,}{\dif \ttau}\delta(\mathcal{W}_m)\,.
\end{equation}
Putting all these together we can finally perform the last integral. The second term in (\ref{inttao}) is trivial due to the presence of the $\delta$-function. The first and third terms can be integrated by parts, though one can easily see that the latter one vanishes. The final result reads
\begin{equation}\label{EAfinal}
\mathcal{E}_{A}=\frac{\sqrt{\lambda}}{8\pi ^{2}}\left[\frac{A_{1}}{|\Xi_{m}|\Xi^{2}}+\frac{1}{|\Xi_{m}|}\frac{\partial}{\partial \ttau}\left(\frac{A_{0}+z_{m}A_{1}}{\Xi_{m}\Xi}\right)\right]\bigg|_{\ttau=\ttau_r}\,.
\end{equation}

 We can repeat the same process to get $\EB$. After the integration over $\dif^4x'$ we get
 \begin{equation}
  \hspace{-0.7cm}
 \EB = \frac{\sqrt{\lambda}}{\pi^2}
 \int \dif \ttau\,\dif z\, \delta'''(\mathcal{W}_q + 2z\Xi)
 [B_0(\ttau) + z B_1(\ttau)+ z^2 B_2(\ttau)],
 \end{equation}
 where the coefficients are given by
 \begin{eqnarray}\label{Bs}
 B_0(\ttau) &=& (x-\tx)^i(x-\tx)_i\left(\tfrac{4}{3}+\tups^j\tups_j\right)
 -\left[(x-\tx)^i\tups_i\right]^2,
 \nonumber \\
 B_1(\ttau) &=& 2(x-\tx)^i(x-\tx)_i\tups^j \ta_j-\tfrac{2}{3}(x-\tx)^i\tups_i\left[4+3(x-\tx)^j \ta_j\right],
 \\
 B_2(\ttau) &=& \tups^i\tups_i\left[\tfrac{4}{3}+2(x-\tx)^j \ta_j\right]-2(x-\tx)^i\tups_i\tups^j \ta_j\nn &&+\,\ta^2\left\{(x-\tx)^i(x-\tx)_i\tups^j\tups_j
 -\left[(x-\tx)^i\tups_i\right]^2\right\}.
 \nonumber
 \end{eqnarray}
 After carrying out the integrals we find
 \begin{eqnarray}\label{EBfinal}
\mathcal{E}_{B}&=&-\frac{\sqrt{\lambda}}{8\pi ^{2}}\left[\frac{B_{2}}{|\Xi_{m}|\Xi^{3}}+\frac{1}{|\Xi_{m}|}\frac{\partial}{\partial \ttau}\left(\frac{B_ {1}+2z_{m}B_{2}}{2\Xi_{m}\Xi^2}\right)\right.
\nn
&&\qquad\quad\left.+\frac{1}{|\Xi_{m}|}\frac{\partial}{\partial \ttau}\left(\frac{1}{\Xi_{m}}\frac{\partial}{\partial \ttau}\left(\frac{B_{0}+z_{m}B_{1}+z_{m}^{2}B_{2}}{2\Xi_{m}\Xi}\right)\right)\right]\bigg|_{\ttau=\ttau_r}\,.
\end{eqnarray}

In the analysis of the rest of this paper, the separation of the energy density into $\EA$ and $\EB$ will play no role.
What does matter physically is a separation into terms without and with $\ttau$-derivatives, which following \cite{HIMT} we denote respectively
\begin{eqnarray}\label{e12}
\cE^{(1)}&=&\frac{\sqrt{\lambda}}{8\pi ^{2}}
\left[\frac{\Xi A_{1}-B_2}{|\Xi_{m}|\Xi^{3}}
\right]_{\ttau=\ttau_r}~,
\\
\cE^{(2)}&=&\frac{\sqrt{\lambda}}{8\pi ^{2}}
\left[\frac{1}{|\Xi_{m}|}
\frac{\partial}{\partial \ttau}\left(\frac{2\Xi A_0-B_1+2z_{m}(\Xi A_1- B_{2})}{2\Xi_{m}\Xi^2}\right)
\right.
\nn
&&\qquad\left.-\frac{1}{|\Xi_{m}|}\frac{\partial}{\partial \ttau}\left(\frac{1}{\Xi_{m}}\frac{\partial}{\partial \ttau}\left(\frac{B_{0}+z_{m}B_{1}+z_{m}^{2}B_{2}}{2\Xi_{m}\Xi}\right)\right)\right]_{\ttau=\ttau_r}~.
\nonumber
\end{eqnarray}
The energy density $\cE=\cE^{(1)}+\cE^{(2)}$ is one of the main results of this paper. As emphasized previously, all tilde variables in (\ref{e12}) are understood to be shorthand for the combinations of physical variables displayed in (\ref{tau2})-(\ref{atilde}).

We learn from  (\ref{e12}) that $\cE^{(1)}$ depends only on $\tups^{\mu}$ and $\ta^{\mu}$, or equivalently, on the physical
velocity $\upsilon^{\mu}$ and external force $\cF^{\mu}$. In contrast with this,
$\cE^{(2)}$ depends on $\tups^{\mu}, \ta^{\mu}, \tj^{\mu}$ and $\tilde{s}^{\mu}$, exactly like the result for $\expec{\trFsq}$ obtained in \cite{trfsq}.
In terms of physical variables, this means that to compute $\cE^{(2)}$ at a particular instant it is not enough to know the velocity of the quark and the force that is applied to it; we also need the first and second derivatives of the force. It should of course be borne in mind that, as mentioned before,
$\cF^{\mu}$ itself always encodes a particular combination of infinitely-many higher derivatives of the quark trajectory---see (\ref{force}).

The most remarkable feature of (\ref{e12}) is that, in spite of its manifestly having been assembled by adding up the backreaction from each infinitesimal segment of the string, or in SYM language, the reemission from the gluonic degrees of freedom at each energy scale, the net result depends only on the quark variables at a single retarded time, much like what happens in the Lienard-Wiechert story of classical electrodynamics. Because of this, disturbances in the gluonic field display no temporal (or, equivalently, radial) broadening. (See however \cite{iancu3}.) As in \cite{HIMT,trfsq}, what has happened is that the retarded structure of the Mikhailov embedding (\ref{mikhsoltilde}) or (\ref{mikhsolzm}) reduces the worldsheet integrand to a total derivative, and the integral then yields a result that can be expressed purely in terms of the behavior of the string endpoint at the latest time that is allowed by causality. The relevant time is determined by condition (\ref{tauretnew}), which reads explicitly
 \begin{equation}\label{tautilderet}
 (x-\tx(\ttau_r))^2=2 z_m(x-\tx(\ttau_r))\cdot \tups(\ttau_r)~,
\end{equation}
and can be rewritten in a much simpler way in terms of the physical variables:
\begin{equation}\label{tauret}
(x-x(\tau_r))^2=-(t-t(\tau_r))^2+(\bx-\bx(\tau_r))^2=-z_m^2.
\end{equation}
This equation describes a two-sheeted hyperboloid about the observation point $x$, which is intersected by the quark worldline twice, once on each sheet. By causality, the root of interest is of course the one in the sheet to the past of $x$, which, in noncovariant notation, corresponds to the retarded time
\begin{equation}\label{tretnc}
t_r=t-\sqrt{(\bx-\bx(t_r))^2+z_m^2}~.
\end{equation}
For $z_m=0$ this sheet of the hyperboloid converges to the past lightcone, but when the quark is finitely-massive, the net propagation of the gluonic disturbances it generates always takes place at a speed slower than that of light.

\section{Remarks on the Total Energy}\label{totalsec}
In order to find the total energy of the system we have to integrate the energy density
over all of space at a fixed observation time,
\begin{equation}
E(t)=\int\dif^3x\,\cE(t,\mathbf{x}).
\end{equation}
It proves convenient to
change the integration variables from $\dif^3x$ to $\dif^3R=R^2\dif R\dif\Omega$,
 where $\bR\equiv\bx-\bx(t_r)\equiv R\,\bn$ denotes the vector from the location of the source to the point of observation.
By writing $\bx=\bR+\bx(\tau_r)$, our first task is then to compute the Jacobian $J=|\partial\bx/\partial\bR|_t$. Using condition (\ref{tauret}), we obtain
\begin{equation}
J=\left|
    \begin{array}{ccc}
      1-\frac{R^1\upsilon^1}{\upsilon^0\sqrt{R^2+z_m^2}} & -\frac{R^2\upsilon^1}{\upsilon^0\sqrt{R^2+z_m^2}} & -\frac{R^3\upsilon^1}{\upsilon^0\sqrt{R^2+z_m^2}} \\
      -\frac{R^1\upsilon^2}{\upsilon^0\sqrt{R^2+z_m^2}} & 1-\frac{R^2\upsilon^2}{\upsilon^0\sqrt{R^2+z_m^2}} & -\frac{R^3\upsilon^2}{\upsilon^0\sqrt{R^2+z_m^2}} \\
      -\frac{R^1\upsilon^3}{\upsilon^0\sqrt{R^2+z_m^2}} & -\frac{R^2\upsilon^3}{\upsilon^0\sqrt{R^2+z_m^2}} & 1-\frac{R^3\upsilon^3}{\upsilon^0\sqrt{R^2+z_m^2}} \\
    \end{array}
  \right|=1-\frac{\bn\cdot\boldsymbol\upsilon}{\upsilon^0\sqrt{1+z_m^2/R^2}}~,
\end{equation}where it is understood that all quantities are evaluated at $\tau=\tau_r$.

We can now reexpress $\dif^3x$ as $J R^2 \dif R \dif\Omega$.
A final useful manipulation is to rewrite the radial integral in terms of the retarded time. Using again (\ref{tauret}), we can deduce that
\begin{equation}
2(t-t_r)\dif t_r+2R\dif R=0\quad\rightarrow\quad \dif R=\sqrt{1+z_m^2/R^2}\dif t_r,
\end{equation}
where $t_r=t(\tau_r)$
is given by (\ref{tretnc}). In terms of proper time we have
\begin{equation}
\dif R=\upsilon^0\sqrt{1+z_m^2/R^2}\dif \tau_r~,
\end{equation}which leads to
\begin{equation}\label{eint}
E(t)=\int\dif \tau_r\dif \Omega\left[1-\frac{\bn\cdot\boldsymbol\upsilon}{\upsilon^0\sqrt{1+z_m^2/R^2}}\right]R^2\upsilon^0\sqrt{1+z_m^2/R^2}\,\cE.
\end{equation}
To process $\cE$, it is helpful to have in hand the following quantities in terms of the physical variables:
\begin{equation}\label{Xiphys}
\Xi=z_m+R\frac{\sqrt{1+z_m^2/R^2}\left(\upsilon^0-z_m^2\bar{\cF}^0\right)-\bn\cdot\left(\boldsymbol\upsilon-z_m^2\,\bar{\!\boldsymbol\cF}\right)}{\sqrt{1-z_m^4\bar{\cF}^2}}~,
\end{equation}\begin{equation}\label{Ximphys}
\Xi_m=R\sqrt{1-z_m^4\bar{\cF}^2}\upsilon^0\sqrt{1+z_m^2/R^2}\left[1-\frac{\bn\cdot\boldsymbol\upsilon}{\upsilon^0\sqrt{1+z_m^2/R^2}}\right]~.
\end{equation}Notice in particular that all terms in the energy density (\ref{e12}) contain a $1/|\Xi_m|$ prefactor, whose angular dependence, seen in (\ref{Ximphys}), conveniently cancels that in the integration measure in (\ref{eint}). No such cancelation would occur if we attempt to carry out the integration in terms of the tilde variables.

We expect the total gauge theory energy $E(t)$ to agree with the total string energy. As reviewed in Section \ref{mikhsec}, the latter was successfully rewritten in \cite{mikhailov,dragtime,lorentzdirac,damping} as the sum of the intrinsic quark energy in (\ref{pq}) and the integrated version of the radiation rate (\ref{radiationrate}),
\begin{equation}\label{stringenergy}
E(t)=\upsilon^{0}\frac{m-{\sqrt{\lambda}\over 2\pi m}
 \bF\cdot\boldsymbol\upsilon}{\sqrt{1-{\lambda\over 4\pi^2 m^4}\cF^2}}
 +\int_{-\infty}^{t-z_m} \!\dif\tau_r\, {\upsilon^0}{\sqrt{\lambda}\, \cF^2 \over 2\pi m^2}
 \left(\frac{1-{\sqrt{\lambda}\over
2\pi m^2} \bF\cdot\boldsymbol\upsilon}{1-{\lambda\over 4\pi^2 m^4}\cF^2}\right)~.
\end{equation}
This is in a form that could be matched by (\ref{eint}) after the angular integration. More specifically,
given the form of (\ref{e12}), it is tempting to conjecture that $\cE^{(1)}$ will directly match the integrand in (\ref{stringenergy}), whereas $\cE^{(2)}$ will end up reducing to a total derivative within the $\tau_r$ integral in (\ref{eint}), and thereby yield the intrinsic quark energy. Alas, this proves to be incorrect.
One obstruction is that the $\ttau$ derivatives in (\ref{e12}) unfortunately cannot be pulled outside the angular integration in (\ref{eint}), because they are meant to be carried out at fixed observation point $x$ rather than at fixed direction $\bn$.
%\footnote{It can be checked that this is not an issue in the infinitely-massive case, $z_m=0$.}
One additionally needs to keep in mind the difference between $\ttau$ and $\tau$ derivatives. And, independently of the way one decides to handle the derivatives, it can be checked explicitly that $\cE^{(1)}$ does \emph{not} reproduce the modified Lienard rate in (\ref{stringenergy}).

The fact that the derivatives in (\ref{e12}) do not commute with the angular integration means that we need to carry them out in full. A very long calculation yields
\begin{eqnarray}\label{elongtildes}
\cE&=&\frac{\sqrt{\lambda}}{8\pi^2}\frac{C_0+z_m C_1+ z_m^2 C_2 + z_m^3 C_3 + z_m^4 C_4}{2|\Xi_m|\Xi_m^4\Xi^3}~,
\\
C_0&=& 2A_1\Xi^5-2B_2\Xi^4+2\dot{A}_0\Xi^5-4A_0\Xi^4\dot{\Xi}-2\dot{B_1}\Xi^4+8B_1\Xi^3\dot{\Xi}-8B_0\Xi^2\dot{\Xi}^2+2B_0\Xi^3\ddot{\Xi}~,
\nonumber\\
C_1&=&6\dot{A}_0\Xi^4\dot{\Xi}-10A_0\Xi^3\dot{\Xi}^2+7B_1\Xi^2\dot{\Xi}^2-7B_0\Xi\dot{\Xi}^3-6B_0\Xi^2\dot{\Xi}\ddot{\Xi}-2A_0\Xi^4\ddot{\Xi}
+6B_1\Xi^3\ddot{\Xi}
\nonumber\\
{}&{}&+2\dot{A}_1\Xi^5+4A_1\Xi^4\dot{\Xi}-2\dot{B}_2\Xi^4-2B_2\Xi^3\dot{\Xi}-\ddot{B}_1\Xi^4+B_0\Xi^3\dddot{\Xi}~,
\nonumber\\
C_2&=&6\dot{A}_0\Xi^3\dot{\Xi}^2-8A_0\Xi^2\dot{\Xi}^3-4A_0\Xi^3\dot{\Xi}\ddot{\Xi}+6\dot{A}_1\Xi^4\dot{\Xi}+2A_1\Xi^3\dot{\Xi}^2-2A_1\Xi^4\ddot{\Xi}
-\dot{B}_2\Xi^3\dot{\Xi}
\nonumber\\
{}&{}&-4B_2\Xi^2\dot{\Xi}^2+4B_2\Xi^3\ddot{\Xi}-2\ddot{B}_1\Xi^3\dot{\Xi}+3\dot{B}_1\Xi^2\dot{\Xi}^2-\ddot{B}_2\Xi^4-B_1\Xi^2\dot{\Xi}\ddot{\Xi}
+B_1\Xi^3\dddot{\Xi}
\nonumber\\
{}&{}&+2B_1\Xi\dot{\Xi}^3-2B_0\Xi\dot{\Xi}^2\ddot{\Xi}+B_0\Xi^2\dot{\Xi}\dddot{\Xi}+3\dot{B}_1\Xi^3\ddot{\Xi}-3B_0\Xi^2\ddot{\Xi}^2
-2B_0\dot{\Xi}^4~,
\nonumber\\
C_3&=&6\dot{A}_1\Xi^3\dot{\Xi}^2-4A_1\Xi^3\dot{\Xi}\ddot{\Xi}+\dot{B}_2\Xi^2\dot{\Xi}^2-2B_2\Xi^2\dot{\Xi}\ddot{\Xi}-2\ddot{B}_2\Xi^3\dot{\Xi}
+B_2\Xi^3\dddot{\Xi}+2\dot{A}_0\Xi^2\dot{\Xi}^3
\nonumber\\
{}&{}&-2A_0\Xi\dot{\Xi}^4-2A_0\Xi^2\dot{\Xi}^2\ddot{\Xi}-B_2\Xi\dot{\Xi}^3-\ddot{B}_1\Xi^2\dot{\Xi}^2+\dot{B}_1\Xi\dot{\Xi}^3
-B_1\Xi\dot{\Xi}^2\ddot{\Xi}+B_1\Xi^2\dot{\Xi}\dddot{\Xi}
\nonumber\\
{}&{}&+3\dot{B}_1\Xi^2\dot{\Xi}\ddot{\Xi}+3\dot{B}_2\Xi^3\ddot{\Xi}-3B_1\Xi^2\ddot{\Xi}^2~,
\nonumber\\
C_4&=&2\dot{A}_1\Xi^2\dot{\Xi}^3-2A_1\Xi^2\dot{\Xi}^2\ddot{\Xi}-\ddot{B}_2\Xi^2\dot{\Xi}^2+B_2\Xi^2\dot{\Xi}\dddot{\Xi}
+3\dot{B}_2\Xi^2\dot{\Xi}\ddot{\Xi}-3B_2\Xi^2\ddot{\Xi}^2~,
\nonumber
\end{eqnarray}
where dots denote derivatives with respect to $\ttau_r$. The complexity of this expression (which is made much worse when writing it out it in terms of the physical variables) unfortunately makes it impossible to carry out the angular integral in (\ref{eint}) for generic trajectories. We are not even allowed to expand in powers of $z_m$ to pursue the leading finite-mass effect, because the expansion parameter would be $z_m/R$, which can be small or large as we integrate over $\tau_r$.  In Section \ref{farsec} we will perform a different comparison between the energies inferred from the worldsheet and bulk approaches. But prior to that, in the following section we will examine the bulk result (\ref{e12}) for a few concrete examples.

\section{A Few Examples}\label{examplessec}

In order to gain some information about which terms in (\ref{e12}) encode the radiative and intrinsic components of the gluonic field, whose integrated values are known from the worldsheet analysis of \cite{dragtime,lorentzdirac,damping}, we here investigate some particular cases.

\subsection{Static quark}\label{staticsubsec}

The static quark is the simplest solution to the equation of motion (\ref{eom}) and from the bulk perspective corresponds to a straight string stretching from $z=z_m$ up to $z=\infty$. The quark data are given by
\begin{equation}
\mathcal{F}^{\mu}=(0,0,0,0),\qquad x^{\mu}=(t_q,0,0,0),\qquad \upsilon^{\mu}=(1,0,0,0),\qquad a^{\mu}=(0,0,0,0),
\end{equation}
with $\tau=t_q$ and $\gamma_q=1$. In view of (\ref{tau2})-(\ref{atilde}), this translates into the auxiliary data
\begin{equation}
\qquad \tx^{\mu}=(t_q-z_m,0,0,0),\qquad \tups^{\mu}=(1,0,0,0),\qquad\ta^{\mu}=(0,0,0,0)
\end{equation}
and $\dif\tilde{\tau}=\dif\tau=\dif t_q$. The coefficients (\ref{As}) and (\ref{Bs}) are then
\begin{equation}
 A_0=3,\qquad A_1=0,\qquad B_0=\tfrac{4}{3}r^2,\qquad B_1=0,\qquad B_2=0,
\end{equation}
where $r=|\mathbf{x}|$. Also, from (\ref{Wq}) and (\ref{Xim}),
\begin{equation}
\Xi=t-t_q+z_m,\qquad \Xi_m=t-t_q.
\end{equation}
Finally, all the expressions have to be evaluated at the retarded time (\ref{tauret}),
\begin{equation}
t_r=t-\sqrt{r^2+z_m^2}.
\end{equation}
Putting all these together we get
\begin{eqnarray}\label{edstatic}
\cE&=&\frac{\sqrt{\lambda}}{8\pi ^{2}}\left[\frac{1}{(t-t_q)}\frac{\partial}{\partial t_q}\left(\frac{3}{(t-t_q)(t-t_q+z_m)}\right)
\right.
\nn
&&\qquad\left.-\frac{1}{(t-t_q)}\frac{\partial}{\partial t_q}\left(\frac{1}{(t-t_q)}\frac{\partial}{\partial t_q}\left(\frac{2r^2}{3(t-t_q)(t-t_q+z_m)}\right)\right)\right]\bigg|_{t_q=t_r}\,,\nonumber\\
&=&\frac{\sqrt{\lambda }}{24 \pi ^2}\frac{3 \left(3 (t-t_q)^2-2 r^2\right) (2 t-2t_q+z_m) (t-t_q+z_m)-4 r^2 (t-t_q)^2}{(t-t_q)^5 (t-t_q+z_m)^3}\bigg|_{t_q=t_r}\,,\\
&=&\frac{\sqrt{\lambda }}{12 \pi ^2r^4}\left(1+\frac{z_m\left(r^2-2z_m^2\right)\left(3r^2+z_m^2\right)}{2\left(r^2+z_m^2\right)^{5/2}}\right)\,.\nonumber
\end{eqnarray}
This shows that the energy density is, as expected, purely near field (for large $r$ it decays as $\sim 1/r^4$), and it comes from the terms $A_0$ and $B_0$ which appear inside $\cE^{(2)}$. For $m\to\infty$ ($z_m\to 0$) this is the Coulombic profile expected for a pointlike charge, with a numerical coefficient that matches previous results \cite{fgmp,HIMT,shuryak}. For finite $m$ the profile is still Coulombic far away from the origin but becomes nonsingular at the location of the quark,
\begin{equation}
\cE=\frac{9 \sqrt{\lambda}}{32 \pi^2 z_m^4}\left(1-\frac{65 r^2}{27 z_m^2}+\mathcal{O}\left(r^4/z_m^4\right)\right).
\end{equation}
Integrating (\ref{edstatic}) over all space we get the total energy of the quark,
\begin{equation}\label{egystatic}
E=4\pi\int dr r^2 \mathcal{E}=\frac{\sqrt{\lambda}}{2\pi z_m}=m,
\end{equation}
which agrees with the worldsheet result (\ref{pq}) for the intrinsic energy, as expected.

It is worthwhile comparing this expression with the known expectation value of the Lagrangian density \cite{martinfsq,trfsq},
\begin{equation}\label{f2static}
\frac{1}{4 g_{YM}^2}\expec{\tr{F^2}+\ldots}={\sqrt{\lambda}\over 16\pi^{2}r^4}
\left(1-\frac{z_m^3\left(z_m^2+{5\over 2}r^2\right)}{\left(r^2+z_m^2\right)^{5/2}}\right)=\frac{15 \sqrt{\lambda }}{128 \pi ^2 z_m^4}\left(1-\frac{7 r^2}{3 z_m^2}+\mathcal{O}\left(r^4/z_m^4\right)\right),
\end{equation}
Recall that
\begin{equation}\label{OpT00}
\frac{1}{4 g_{YM}^2}\expec{\tr{F^2}+\ldots}=\frac{1}{2 g_{YM}^2}\tr{(E^2-B^2)}+\cdots~,\qquad\expec{T_{00}}=\frac{1}{2 g_{YM}^2}
\mathrm{Tr}\, (E^2+B^2)+\cdots~,
\end{equation}
where the dots represent contributions from the other SYM fields. If the chromomagnetic field is not present, as in the case of the static quark, then the difference in the two expectation values implies that significant contributions are received from these additional fields.

\begin{figure}[ht]
$$
\begin{array}{cc}
  \includegraphics[width=7cm]{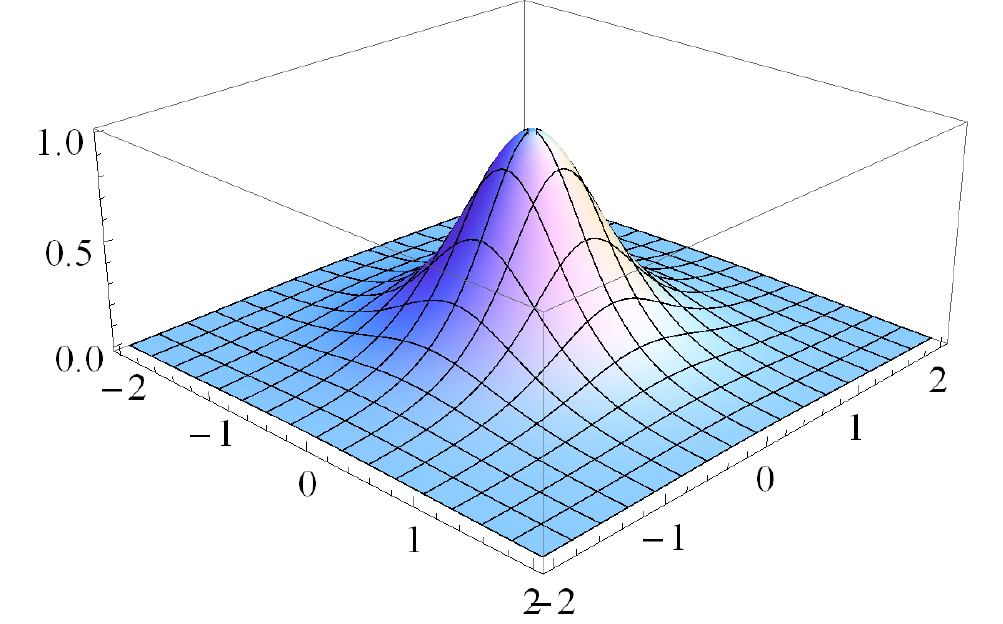} & \includegraphics[width=7cm]{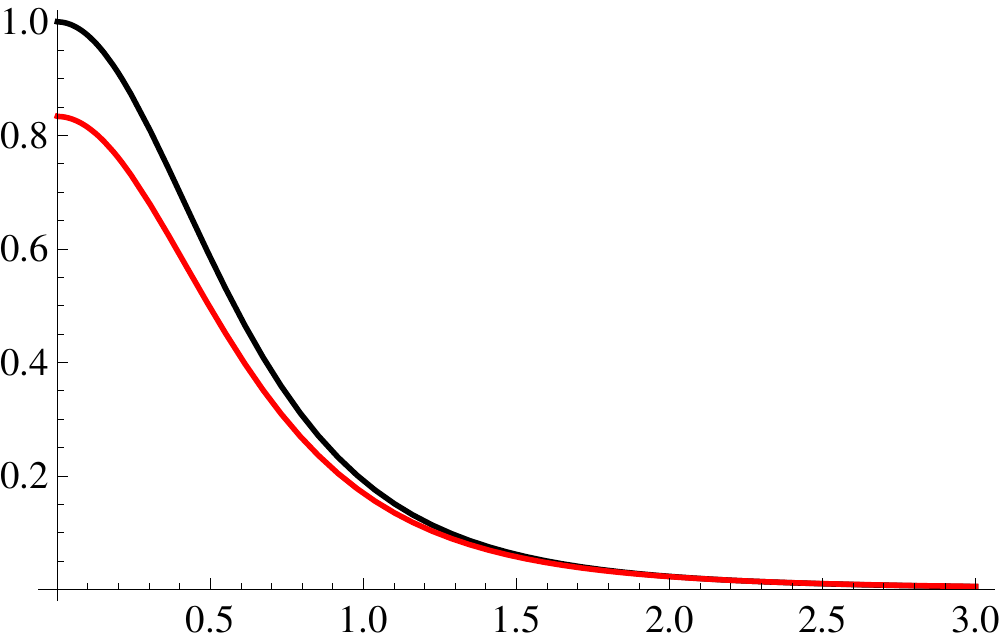}
\end{array}
$$
\begin{picture}(0,0)
 \put(0,143){\small $\cE$,{\color{red}$\expec{\tr{F^2}}$}}
 \put(204,21.5){\small $r$}
 \put(-188.5,117.5){\small $\cE$}
 \put(-151,32){\small $x$}
 \put(-45,33){\small $y$}
 \end{picture}
\caption{Left panel: energy density $\cE$ sourced by a static quark with finite mass, $z_m=1$, normalized such that the value of the profile at the position of the quark is unity. Right panel: comparison of the Lagrangian density (red) sourced by a static quark with respect to its energy density $\cE$ (black), normalized such that the spatial integrals are the same. \label{figstatic1}}
\end{figure}
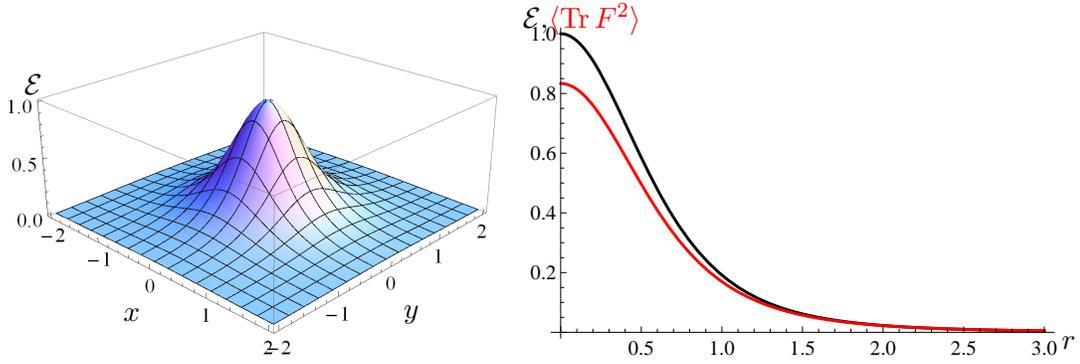

In the left panel of Figure~\ref{figstatic1} we plot the expectation value of the energy density given in (\ref{edstatic}). The profile is in close resemblance to that of the field strength (\ref{f2static}). Perhaps the only noticeable difference is the amplitude, but the shape itself also varies slightly. To make this explicit we show in the right panel a comparison in which we rescale the numerical coefficient of (\ref{f2static}) such that its integral is equal to the energy (\ref{egystatic}). The comparison shows that the outer regions contribute more significantly to the total Lagrangian than to the total energy.

\subsection{Constant velocity}\label{constantsubsec}

We will now examine a uniformly translating quark, which is the other trivial solution to the  equation of motion (\ref{eom}) when the force is zero. This is, of course, related by a boost to the static case, but it is useful to work it out explicitly to track the coefficients that contribute. For motion along the $x_{\longi}\equiv x_3$ axis, the expressions are
\begin{equation}
\mathcal{F}^{\mu}=(0,0,0,0),\qquad x^{\mu}=(t_q,0,0,t_q \upsilon_q),\qquad \upsilon^{\mu}=(\gamma_q,0,0,\gamma_q \upsilon_q),\qquad a^{\mu}=(0,0,0,0).
\end{equation}
Using the above one gets
\begin{equation}
\qquad \tx^{\mu}=(t_q-z_m\gamma_q,0,0,(t_q-z_m\gamma_q) \upsilon_q),\qquad \tv^{\mu}=(\gamma_q,0,0,\gamma_q \upsilon_q),\qquad\ta^{\mu}=(0,0,0,0),
\end{equation}
and $\dif\tilde{\tau}=\dif\tau=\dif t_q/\gamma_q$. The coefficients (\ref{As}) and (\ref{Bs}) are
\begin{eqnarray}
A_0=1+2\gamma_q^2,\qquad
A_1=0,\qquad
B_0=(\tfrac{4}{3}+\gamma_q^2 \upsilon_q^2)x_{\perp}^2+\tfrac{4}{3}[x_{\longi}-(t_q-z_m \gamma_q) \upsilon_q]^2,\\
B_1=-\tfrac{8}{3}[x_{\longi}-(t_q-z_m \gamma_q) \upsilon_q]\gamma_q \upsilon_q,\qquad
B_2=\tfrac{4}{3}\gamma_q^2 \upsilon_q^2.\qquad\qquad\qquad\quad \nonumber
\end{eqnarray}
We also have
\begin{equation}
\Xi=(t-t_q)\gamma_q-(x_{\longi}-t_q \upsilon_q)\gamma_q \upsilon_q+z_m,\qquad
\Xi_m=(t-t_q)\gamma_q-(x_{\longi}-t_q \upsilon_q)\gamma_q \upsilon_q.
\end{equation}
Finally, the condition for the retarded time (\ref{tauret}) reads
\begin{equation}
t_r=\gamma_q\left(\gamma_q(t-x_{\longi} \upsilon_q)- \sqrt{\gamma_q^2(x_{\longi}-t \ups_q)^2+ x_{\perp}^2+ z_m^2}\,\right).
\end{equation}
Putting all these together we get
\begin{eqnarray}\label{edvelocity}
\cE&=&\frac{\gamma_q^2 \sqrt{\lambda }}{48 \pi ^2 (z_m+\sqrt{x_{\longi}'^2+x_\perp^2+z_m^2})^3}\left(\frac{3 z_m^2(\ups_q^2 x_\perp^2+4 z_m^2)}{(x_{\longi}'^2+x_\perp^2+z_m^2)^{5/2}}+\frac{9 z_m (\upsilon_q^2 x_\perp^2+4 z_m^2)}{(x_{\longi}'^2+x_\perp^2+z_m^2)^2}
\right.\\
&&\qquad\qquad\left.+\frac{8 \upsilon_q^2 x_\perp^2+2z_m^2(19-\upsilon_q^2)}{(x_{\longi}'^2+x_\perp^2+z_m^2)^{3/2}}+\frac{6 z_m(3-\upsilon_q^2)}{x_{\longi}'^2+x_\perp^2+z_m^2}+\frac{4(1-\upsilon_q^2)}{(x_{\longi}'^2+x_\perp^2+z_m^2)^{1/2}}\right),
\nonumber
\end{eqnarray}
where for compactness we have employed $x_{\longi}'\equiv\gamma_q(x_{\longi}-t \upsilon_q)$.

The large distance behavior of the above expression is dominated by the last term. As in the static case, one finds that it decays as $\sim 1/r'^4$, with $r'^2=x_{\longi}'^2+x_\perp^2$, so the energy density is again just near field, as expected. {}From the calculation we see that (\ref{edvelocity}) comes from the terms $A_0$, $B_0$, $B_1$ and $B_2$. This suggests that the term $A_1$ might entirely encode radiation, given that this is the most general example in which one does not expect the quark to radiate.

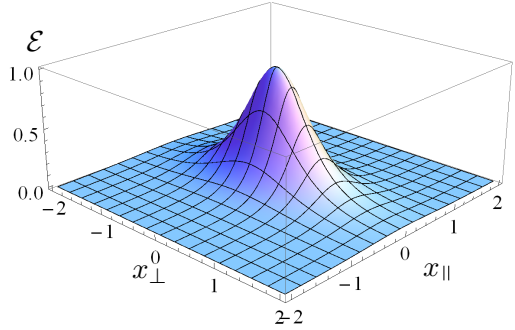
\begin{figure}[ht]
  \includegraphics[width=7cm]{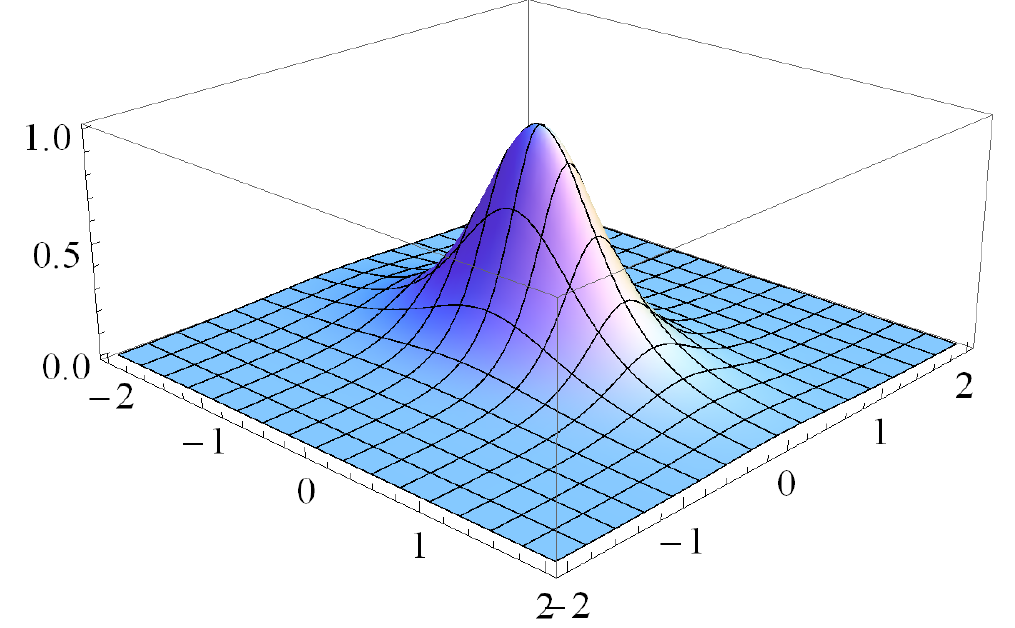}
\begin{picture}(0,0)
 \put(-192,105){\small $\cE$}
 \put(-152,19){\small $x_{\perp}$}
 \put(-42,20){\small $x_{\longi}$}
 \end{picture}
\caption{Energy density $\cE$ of a quark moving with constant velocity $\ups=0.8$ at time $t=0$. The horizontal axes are in units of $z_m = 1$, and the vertical axis is normalized such that the profile at the position of the quark is unity. The pattern obtained displays the expected Lorentz contraction in the transverse direction. For different times the plot is of course shifted in the direction of the motion of the quark.\label{figstatic3}}
\end{figure}

In Figure~\ref{figstatic3} we plot the expectation value of the energy density given in (\ref{edvelocity}) for a quark moving with constant velocity. We chose a relatively large value for the velocity, $\ups=0.8$, in order to have a noticeable Lorentz contraction effect.

To obtain the integrated energy we first decompose the transverse directions in cylindrical coordinates. Then, $\dif^3x=2\pi x_\perp \dif x_\perp\dif x_{\longi}=2\pi x_\perp\dif x_\perp\dif x_{\longi}'/\gamma_q$, and using (\ref{edvelocity}),
\begin{equation}
E=\frac{2\pi}{\gamma_q}\int_{-\infty}^{\infty}\dif x_{\longi}'\int_0^{\infty} \dif x_\perp x_\perp\cE(t,\mathbf{x})=\frac{\gamma_q  \sqrt{\lambda }}{12 \pi }\int_{-\infty}^{\infty}\frac{\dif z'}{z'^2}\left(1+\frac{z_m(z'^2-z_m^2)}{(z'^2+z_m^2)^{3/2}}\right)=\frac{\gamma_q\sqrt{\lambda}}{2\pi z_m}=\gamma_q m,
\end{equation}
which is the relativistic intrinsic energy for a moving particle.

\subsection{No forced rest}\label{forcedrestsubsec}

For our next example, we would like to find a situation where the quark emits gluonic radiation, which will happen whenever it is externally forced. As described below (\ref{e12}), our general expression for the energy density depends not only on the physical quark trajectory $x^{\mu}(\tau)$ but also on the applied force $\cF^{\mu}(\tau)$. These functions, of course, are related by the equation of motion (\ref{eom}). In \cite{damping} it was emphasized that this equation implies a definite quark motion for any given applied force (and choice of initial conditions), which means in particular that, in contrast with the original Lorentz-Dirac equation, it has no self-accelerating solutions. But it was also observed that, curiously, the converse is not true, in that (\ref{eom}) amounts to a \emph{differential} equation for the external force, and therefore any given quark trajectory can be associated with a one-parameter family of possible force functions. This is related with the fact that the finite-size gluonic cloud of the quark can be distorted dynamically.

A specific example discussed in \cite{damping} is the case where the quark is static (or translating uniformly), where (\ref{eom}) was found to admit the general solution
\begin{equation}\label{forcedrest}
F(t)=\frac{\sqrt{\lambda}}{2\pi z_m^2}\mathrm{sech}\!\left(\frac{t-t_0}{z_m}\right)~,
\end{equation}
with $t_0$ an integration constant. Only for $t_0\to\pm\infty$ does one recover the
straightforward result $F(t)=0$. The other solutions describe situations where the effect of the force is not to accelerate the quark but to distort it. This distortion translates into a time-dependence both in its intrinsic and its radiated energy. This forced rest scenario would thus appear to be the simplest possible trajectory for which one could explore the radiation profile.

We have found, however, that a sign was missed in \cite{damping} which implies that the force function (\ref{forcedrest}) actually solves (\ref{eom}) only for $t\ge t_0$. One must then paste this valid portion of (\ref{forcedrest}) onto an appropriate solution of the equation of motion for
$t\le t_0$. By continuity (of both the force and the associated string profile (\ref{mikhsolzm})), the desired completion must also satisfy $F(t_0)=\sqrt{\lambda}/2\pi z_m^2$, and it is easy to see that, if we insist on having the quark permanently at rest, the unique solution to (\ref{eom}) with this final condition is the constant $F(t)=\sqrt{\lambda}/2\pi z_m^2$ for all $t\le t_0$.
This is exactly the critical value (\ref{Fcrit}) of the force, at which both the intrinsic and the radiated energy of the quark diverge, so our setup becomes unphysical. In the bulk, the issue is that the extreme value of the force for $t\le t_0$ causes the string to become completely parallel to the AdS boundary, which is why it has infinite energy. The only way to avoid this unphysical regime is to take $t_0\to-\infty$, but in this limit the force (\ref{forcedrest}) just vanishes.

So, in the case of the static (or uniformly translating) quark, contrary to what was believed in \cite{damping} there is in the end only one physical choice for the force. In the three cases with acceleration that were analyzed numerically in \cite{trfsq}, the various force profiles allowed for a given quark motion were found to differ only by some transient behaviour, which becomes irrelevant if the trajectory is really known for all time and the initial condition is imposed in the remote past.
  This is similar to what we have found here for (\ref{forcedrest}). We believe that it is true in general that ultimately there is always only one physically sensible solution. Certainly, if we leave out of the discussion situations where the force becomes critical, then we have already noted in Section \ref{mikhsec} that the equation of motion can be transcribed into the form (\ref{force}), where a choice of quark worldline completely determines the applied force.

\subsection{Uniform circular motion}\label{circularsubsec}

We will now consider the case of uniform circular motion. This is a classic example, that has been studied in detail for the case of an infinitely-massive quark at zero temperature \cite{liusynchrotron,veronika} and in the presence of a thermal plasma \cite{Chesler:2011nc}. The surprising feature of this configuration is that
the radiation pattern was found to coincide with the synchrotron spiral familiar from classical electrodynamics, giving the first indication of the absence of temporal/radial broadening emphasized in \cite{iancu1,HIMT,trfsq}.

The trajectory of a quark moving with constant angular velocity $\omega$ in a circle of radius $\rho_m$ can be described by
\begin{equation}\label{rotatrajectory}
x^{\mu}=(t, \rho_m\cos{\omega t},\rho_m\sin{\omega t},0),
\end{equation}
with corresponding four-velocity
\begin{equation}\label{velocity}
\upsilon^{\mu}=(\gamma, -\gamma \omega \rho_m\sin{\omega t},\gamma\omega \rho_m\cos{\omega t},0)~,\quad \text{with}\quad \gamma=\frac{1}{\sqrt{1-\omega^2 \rho_m^2}}~.
\end{equation}

In order to find the energy density sourced by a quark following this trajectory we first have to solve the generalized Lorentz-Dirac equation (\ref{eom}) in order to find the external force $\mathcal{F}^{\mu}$ needed to sustain such kind of motion. This is in general a challenging task. However, for the case under consideration we can use the following trick, discussed originally in \cite{trfsq}.
The string embedding dual to an infinitely massive quark undergoing constant circular motion is known to take the form of a uniformly
rotating spiral \cite{liusynchrotron}. In the case of finite mass, then, in order to satisfy the requirement that the physical string endpoint
at $z_m$ undergo uniform circular motion, we just have to impose that the fictitious endpoint at $z=0$ also rotate uniformly with the same frequency (and with an appropriately shifted phase). We will then be able to make direct use of the formula (\ref{e12}) for the energy density in terms of the auxiliary tilde variables.

In the static gauge, the string embedding for such an spiral is described as $X^m=(t,\vec{r}(t,z),z)$ where, in spherical coordinates $\{r,\theta,\varphi\}$, the three-vector $\vec{r}$ is given by
\begin{equation}
\vec{r}(t,z)=(\rho(z),\tfrac{\pi}{2},\phi(z)+\omega t)~.
\end{equation}
The embedding functions $\rho(z)$ and $\phi(z)$ read \cite{liusynchrotron}
\begin{equation}
\rho(z)=\rho_0\sqrt{1+\frac{\omega^2 z^2}{1-\omega^2\rho_0^2}}\quad\text{and}\quad\phi(z)=-\frac{\omega z}{\sqrt{1-\omega^2 \rho_0^2}}+\arctan \left(\frac{\omega z}{\sqrt{1-\omega^2 \rho_0^2}}\right)~,
\end{equation}
respectively. From the above, it is clear that we must choose the  $\rho_0$ radius at the fictitious boundary endpoint such that
\begin{equation}\label{rotatingradius}
\rho(z_m)=\rho_0\sqrt{1+\frac{\omega^2 z_m^2}{1-\omega^2\rho_0^2}}=\rho_m~.
\end{equation}
And if we truly want $\varphi(0,z_m)=0$ we must change the time origin according to $t\to t-t_s$, with
\begin{equation}
t_s=-\frac{z_m}{\sqrt{1-\omega^2 \rho_0^2}}+\frac{1}{\omega}\arctan \left(\frac{\omega z_m}{\sqrt{1-\omega^2 \rho_0^2}}\right)
\end{equation}
(or alternatively, the same effect can be accomplished by an appropriate shift on the phase $\varphi$).

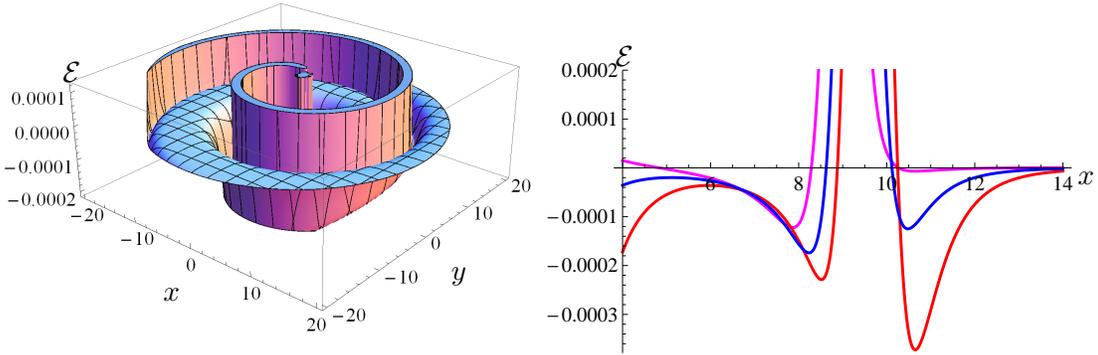
\begin{figure}[ht]
$$
\begin{array}{cc}
  \includegraphics[width=7cm]{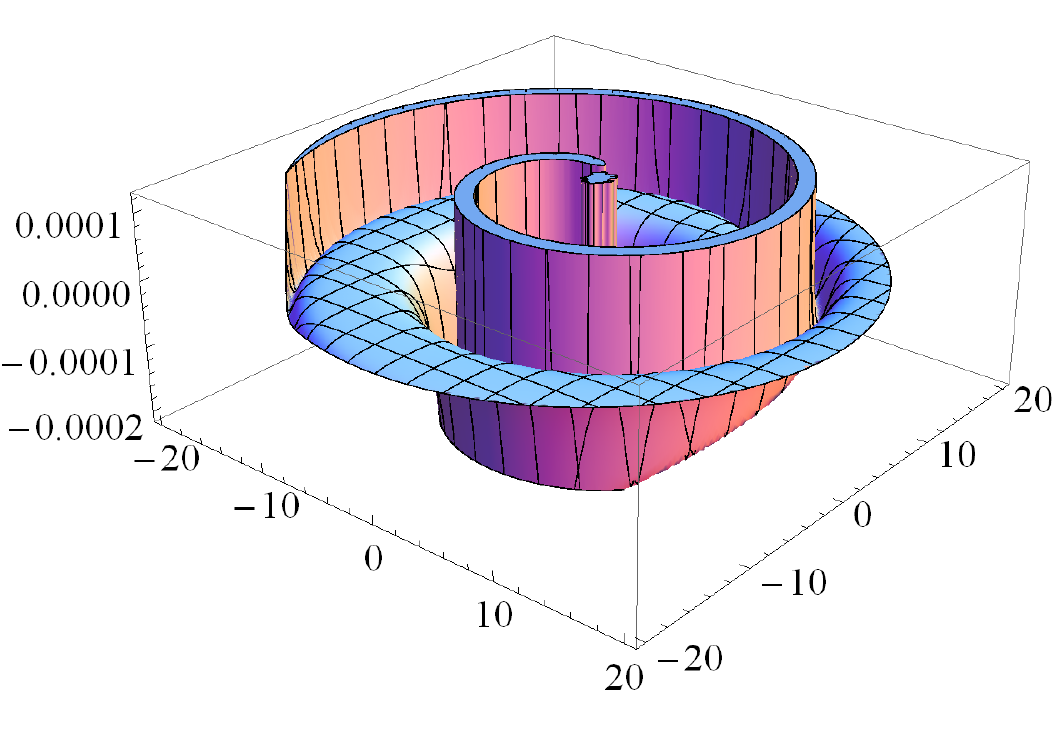} & \includegraphics[width=7cm]{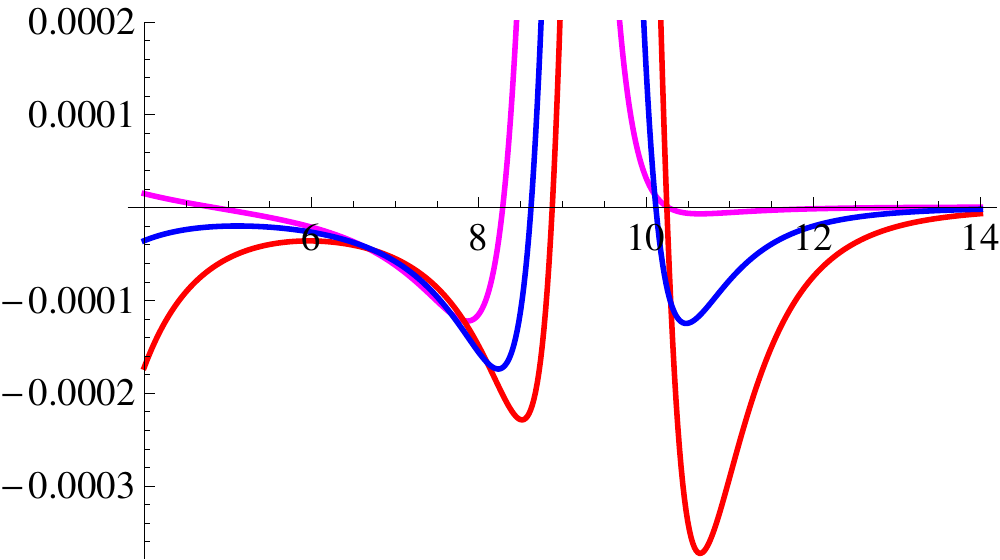}
\end{array}
$$
\begin{picture}(0,0)
 \put(29,126){\small $\cE$}
 \put(204,81){\small $x$}
 \put(-180,120){\small $\cE$}
 \put(-142,37){\small $x$}
 \put(-33,45){\small $y$}
\end{picture}
\caption{Left panel: energy density sourced by a heavy quark undergoing uniform circular motion in the $xy$-plane, with parameters $\lambda=1$, $z_m=5\times10^{-2}$, $\omega=1/2$ and $\rho_m=1$ (for which (\ref{rotatingradius}) yields $\rho_0\simeq0.874$). The finite but proportionately large field of the spiral pattern is capped off in order for the vertical axis to remain at a scale where the details of the waves are clearly visible. The radiation pattern displays no broadening, similar to the case of an infinitely-massive quark \cite{liusynchrotron,veronika,iancu1}, but the effect of negative energy densities is reduced. Right panel: details of the energy density at $\theta=\pi/2$ and $\varphi=0$ for various values of $z_m=10^{-1}$ (red), $5\times10^{-1}$ (blue) and 1 (magenta).\label{circularfigs}}
\end{figure}

To derive explicit results for the energy density we additionally need to determine the retarded time $t_r$ from (\ref{tretnc}). This equation turns out to be transcendental, so we must proceed numerically. The final result for $\cE$ is shown in Figure~\ref{circularfigs} (left panel), for a sample choice of parameter values. The spiral profile of the gluonic fields is clearly in close resemblance with the synchrotron pattern for an infinitely massive quark computed in \cite{liusynchrotron,veronika,iancu1}. At finite mass, however, we have a modified pattern with somewhat different width of the spiral arm and a net subluminal propagation speed, as dictated by (\ref{tretnc}). This effect becomes manifest in the right panel of
Figure~\ref{circularfigs}: as $z_m$ is increased, the pulse moves slightly to the left, indicating that in the same time interval the disturbance has traveled less distance. For the parameters chosen in the plot, we observe that the the position of the spiral along the $x$-axis shifts back by $\sim10\%$ as we increase $z_m$ from zero to unity (larger values would make the size of the quark larger than the radius of the circle). In the same interval, the height of the peak decreases in a roughly linear fashion by about $45\%$, while the width of the spiral increases somewhat, almost linearly, by $\sim20\%$. This behavior is in agreement with what we expect from the usual UV/IR connection in AdS/CFT: our bulk scale $z_m$ maps into a length scale $\ell\sim z_m$ in the CFT, which in the present context can be identified as the characteristic width of the gluonic cloud. It is then natural that any length scale in our problem gets rescaled by a factor of order $z_m$.

Another feature of the radiation pattern that is worth emphasizing is that, even in the limit $z_m\to 0$, for large enough velocity the energy density is negative in the regions of space directly ahead of and behind the spiral. This property was first observed in \cite{liusynchrotron} and reflects the fact that, via holography, our result for the expectation value of the energy density is fully quantum mechanical. Indeed, the energy density in a quantum field theory need
not be positive everywhere; only its integral over all space is constrained to be positive (and bounded by quantum inequalities---see \cite{NegativeEDinQFT} for references).\footnote{To our knowledge, negative energy densities were first discussed  in the AdS/CFT context in \cite{Polchinski:1999yd}, where they were argued to arise naturally as a result of the UV/IR connection for processes occuring deep in the bulk of AdS, embodied in a class of nonlocal hidden degrees of freedom of the theory.}
For finite quark mass, our result shows that the amplitudes of the negative-energy pulses are reduced as $z_m$ increases, or equivalently, as the mass of the quark decreases---see the right panel of Figure~\ref{circularfigs}. For the parameters chosen in the plot, we observe that the height of the negative pulse behind the spiral arm is decreased by $\sim50\%$ as we increase $z_m$ from zero to unity, whereas the one of the negative pulse ahead of the spiral decreases by $\sim99\%$. In the same range the (positive) height of the spiral arm itself decreases only by $\sim45\%$. These results appear to be in concordance with the Quantum Interest conjecture \cite{NegativeEDinQFT}. In short, this conjecture states that a positive energy pulse must overcompensate the negative energy pulse by an amount which is a monotonically increasing function of the pulse separation. In our setting, as the size $z_m$ of the quark decreases, the positive-energy pulse increases in amplitude and becomes more localized, so it can (over-)compensate for the presence of a larger pulse with negative energy.

\subsection{Harmonic motion}\label{harmonicsubsec}

Let us now center our attention in another non-trivial example in which we expect radiation to be emitted: the prototypical case of a quark undergoing one-dimensional harmonic motion. Without loss of generality, we can choose $x\equiv x^1$ as the direction of oscillation. In this case, the trajectory is given by
\begin{equation}\label{xosc}
x^{\mu}=(t, A \sin(\omega t),0,0)~,
\end{equation}
with corresponding four-velocity
\begin{equation}\label{vosc}
\upsilon^{\mu}=(\gamma, \gamma \omega A \cos(\omega t),0,0)~,\quad \text{where}\quad
\gamma=\frac{1}{\sqrt{1-\omega^2A^2\cos^2(\omega t)}}~.
\end{equation}
For small amplitudes, $\omega^2A^2\ll1$, and in the limit of an infinitely heavy quark, the radiation pattern of such configuration was first studied in \cite{Maeda:2007be} within the linearized approximation. Additionally, various properties of the Lagrangian density (a purely near-field observable) were explored in \cite{cg,trfsq}, both for small and large amplitudes of the motion.

As usual, our finite-mass expression for the energy density (\ref{e12}) requires, apart from the quark data (position, velocity, acceleration and jerk), knowledge of the total external force applied to it, together with its first and second derivatives (sometimes referred to as yank and  tug). More specifically, in the language of (\ref{e12}), we require to determine the auxiliary data of the fictitious string endpoint which is encoded by the tilde variables (\ref{xtilde})-(\ref{atilde}) and their time derivatives. In contrast with the previous subsection, here we do not know the form that the string embedding (\ref{mikhsolzm}) corresponding to harmonic motion takes in static gauge, so we do need to solve the quark equation of motion (\ref{eom}) to find the force that gives rise to our desired trajectory. For one-dimensional harmonic motion this equation demands that $F\equiv F^1$ satisfy
\begin{equation}\label{lorentzdiracoscillate}
\frac{d\bar{F}}{dt}=-\sqrt{1-\omega^2A^2\cos^2(\omega t)}\sqrt{1-z_m^4\bar{F}^2}{\bar{F}\over z_m}
-\frac{\omega^2 A \sin(\omega t)}{1-\omega^2A^2\cos^2(\omega t)}
\left(\frac{1-z_m^4\bar{F}^2}{z_m^2}\right)~.
\end{equation}
Notice that the nonlinearity of this equation implies that the force will be in general non-harmonic, as a consequence both of the extended nature of the quark and of the damping effect due to the emitted radiation.

To solve the above equation one must impose an initial condition for the force at some time $t_0$. This would generate a family of force functions $F(t)$ giving rise to the desired motion for $t>t_0$. As mentioned at the end of Section \ref{forcedrestsubsec}, the different solutions are found numerically to differ only by some transient \cite{trfsq} that disappears at some $(t-t_0)\sim\mathcal{O}(z_m)$. Physically, (\ref{force}) tells us that the initial condition supplied to the numerical integration of (\ref{lorentzdiracoscillate}) could in fact be deduced from the behavior of the quark at previous times. Thus, if we truly want to enforce a certain trajectory for the quark at \emph{all times}, then we are automatically forced to work in the late-time regime where there is no transient.\footnote{In practice this can be implemented easily: we can just solve numerically (\ref{lorentzdiracoscillate}) giving an initial condition at some $t_0<<t_{\text{start}}$, where $t_{\text{start}}$ is the starting time of our simulation. Alternatively, we could set $t_0=t_{\text{start}}$ and fine-tune the initial condition so that there is no transient at $(t-t_0)\ll z_m$.} This mechanism selects a unique physical solution valid for all times.

\begin{figure}[ht]
$$
\begin{array}{cc}
  \includegraphics[width=7cm]{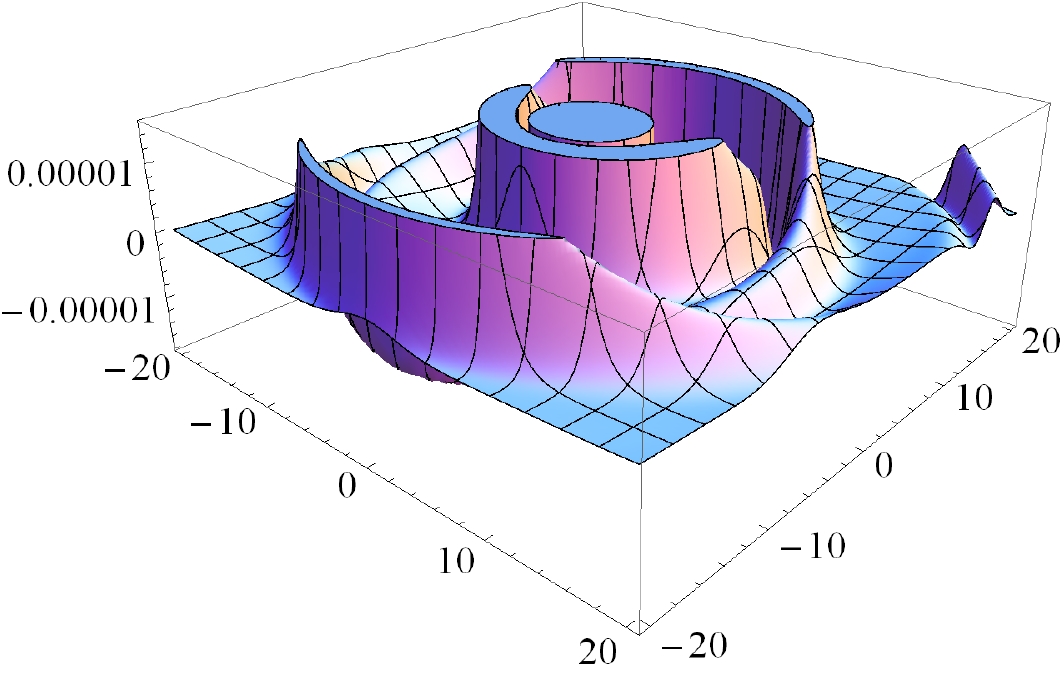} & \includegraphics[width=7cm]{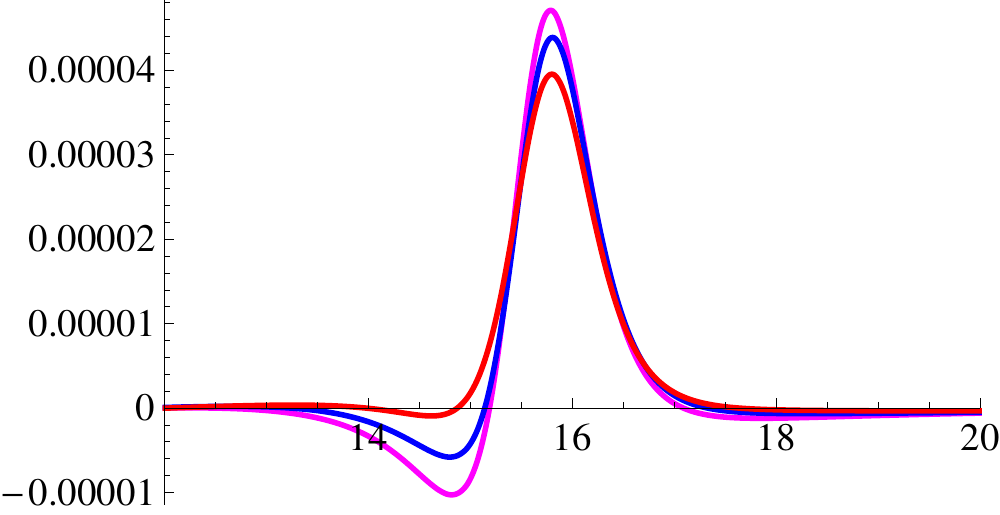}
\end{array}
$$
\begin{picture}(0,0)
 \put(33,118){\small $\mathcal{E}$}
 \put(203,32){\small $x$}
 \put(-178,120){\small $\mathcal{E}$}
 \put(-146,39){\small $x$}
 \put(-32,44){\small $y$}
\end{picture}
\caption{Left panel: energy density sourced by a heavy quark undergoing harmonic motion in the $x$-direction, with parameters $\lambda=1$, $z_m=1$, $\omega=1/2$ and $A=1$. The observation time was set to $t=0$. The finite but proportionately large field of the wavefronts is capped off in order for the vertical axis to remain at a scale where the details of the waves are clearly visible. Right panel: details of the waves at $x2=x3=0$ for various values of $z_m=4\times10^{-1}$ (red), $7\times10^{-1}$ (blue) and 1 (magenta).\label{harmonicfigs}}
\end{figure}
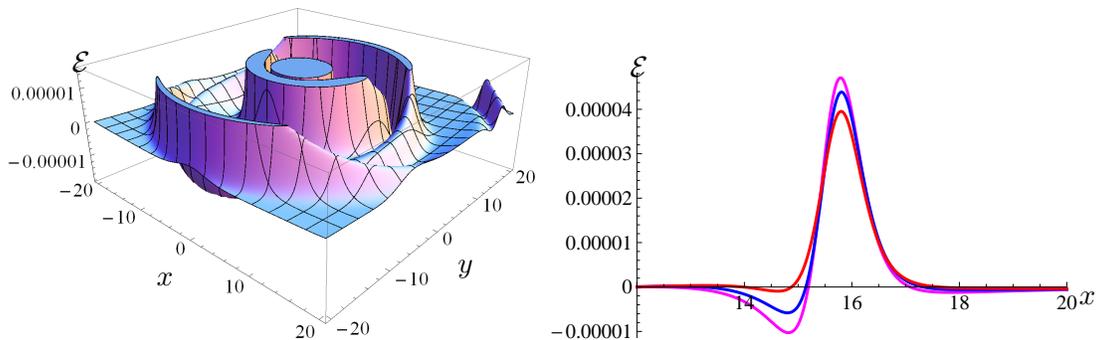

In order to compute the energy density we also need the retarded time $t_r$ which is obtained from (\ref{tretnc}). As in the previous example, this equation turns out to be transcendental, so we solve it numerically. Plugging the results from both numerical integrations in (\ref{e12}) we find the pattern displayed in figure \ref{harmonicfigs}. In the left panel we show the full energy density for sample parameters $\lambda=1$, $z_m=1$, $\omega=1/2$ and $A=1$, for a particular observation time set to $t=0$. The oscillatory motion generates a nonlinear wave, with crests splitting off from the quark every half-cycle.

Some of the features of the energy density resemble those encountered in the previous section for uniform circular motion. In particular, there exist negative-energy regions directly ahead of and behind the wavefronts. As argued before, this result reflects the fact that our computation is fully quantum mechanical. However, there are some important differences regarding the qualitative dependence with respect to $z_m$ that deserve some attention. These differences become evident in the right panel of figure \ref{harmonicfigs}, where we plot details of one of the wavefronts for $\lambda=1$, $\omega=1/2$, $A=1$ and different values of $z_m$. The first one is that the value of $z_m$ now seems to have a negligible effect on the position of the wave. This might seem in conflict with the notion of subluminal propagation speed dictated by (\ref{tretnc}), but the explanation is mundane: while the position of the wavefront plotted in figure \ref{harmonicfigs} is of order $R=|x-x_q|\sim16$, almost twice as large as the position of the spiral studied in figure \ref{circularfigs}, the values of $z_m$ we explored are at most of order unity. And the $z_m$-dependence in (\ref{tretnc}) does become negligible for $R\gg z_m$, as the past hyperboloid approaches the past lightcone. The other two noticeable differences are that, for the case of harmonic motion, the amplitude of the wavefronts increases as we increase the value of $z_m$ and, at the same time, they become more localized. The combination of both effects, the increase in the amplitude and the decrease of the width, seem also in concordance with the Quantum Interest conjecture \cite{NegativeEDinQFT} mentioned already in section \ref{circularsubsec}. Nonetheless, it is worth emphasizing that the central peak (the one just above the position of the quark) does behave as expected: as we increase the value of $z_m$ the amplitude decreases and the width of the gluon cloud increases.

%Problem with UV/IR?

\section{Far-zone Energy and Power} \label{farsec}

Our results for the energy density simplify considerably if we examine them at distances much larger than the quark Compton wavelength,
$R\gg z_m$.
 More specifically, in this section we will consider the limit $z_m/R\to 0$ at fixed emission time
 $t_r$ (equivalently, fixed $\ti_r$),
 so we need to scale the observation time according to $t\simeq t_r+R$.
 {}From (\ref{Wq}), (\ref{As}), (\ref{Xim}), (\ref{tauret}) and (\ref{Bs}) it follows that
$\btR \equiv \bx-\btx(\ti_r)\simeq \bR$ (i.e., $\tR\simeq R$, $\btn\simeq\bn$),
\begin{eqnarray}\label{largeR}
\Xi&\simeq& R(\tv^{0}-\bn\cdot\tilde{\boldsymbol\upsilon})~, \nonumber\\
\Xi_m&\simeq& \Xi+R z_m(\ta^{0}-\bn\cdot\bta)~, \nonumber\\
A_0&\simeq& R (\ta^{0}+\bn\cdot\bta)~, \nonumber\\
A_1&\simeq& R \ta^2(\tv^{0}+\bn\cdot\tilde{\boldsymbol\upsilon})~, \\
B_0&\simeq& R^2\left[\frac{4}{3}+\btv^2-(\bn\cdot\tilde{\boldsymbol\upsilon})^2\right]~,\nonumber\\
B_1&\simeq& R^2 2(\btv\cdot\bta-\bn\cdot\tilde{\boldsymbol\upsilon}\,\bn\cdot\bta)~, \nonumber\\
B_2&\simeq& R^2\ta^2\left[\tilde{\boldsymbol\upsilon}^2-(\bn\cdot\tilde{\boldsymbol\upsilon})^2\right]~. \nonumber
\end{eqnarray}
Employing this in (\ref{e12}), the no-derivative portion of the energy density $\cE\equiv\expec{T^{00}}$ is found to reduce to
\begin{equation}\label{e1rad}
\cE^{(1)}_{\mbox{\scriptsize far}}=\frac{\sqrt{\lambda}}{8\pi^2}
\frac{\ta^2}{R^2(\tups^{0}-\bn\cdot\tilde{\boldsymbol\upsilon})^3\left[\tups^{0}
-\bn\cdot\tilde{\boldsymbol\upsilon}+z_m(\ta^{0}-\bn\cdot\bta)\right]}~.
\end{equation}

The preceding expression has the expected $1/R^2$ falloff expected at large distances, normally taken to be the defining characteristic of radiation.
Following \cite{HIMT}, we can process it a bit further to
obtain the corresponding contribution to the power radiated by the quark. The first step is to exploit the fact that
$\p_t\expec{T^{00}}=-\p_i\expec{T^{i0}}$ and
$\expec{T^{i0}_{\mbox{\scriptsize far}}}\simeq n^i \expec{T^{00}_{\mbox{\scriptsize far}}}$ to rewrite
 \begin{eqnarray}\label{dEdt}
 \frac{dE}{dt}&\equiv& \int d^3 x\,\p_t\cE(t,\bx)=- \int d^3 x\,\p_i\expec{T^{i0}(t,\bx)}\\
 {}&\simeq& -\lim_{R\to\infty}R^2\int d\Omega\,\expec{T^{00}_{\mbox{\scriptsize far}}(t,\bx)}
 =-\lim_{R\to\infty} R^2\int d\Omega\,\cE_{\mbox{\scriptsize far}}(t,\bx)~,\nonumber
 \end{eqnarray}
 where in the second-to-last step we have approximated $\bx\simeq\bR$. This is valid in the long-distance limit we are
  considering, as long as all acceleration on the part of the
 quark takes place within a bounded neighborhood of the origin. The second step is to notice from (\ref{tretnc}), (\ref{Ximphys}) and (\ref{largeR}) that at fixed $\bn$ we have
 \begin{equation}\label{dtdtr}
 \frac{dt}{dt_r}=1-\frac{\bn\cdot\boldsymbol\upsilon}{\upsilon^0}\simeq\frac{\Xi_m}{R\sqrt{1-z_m^4\cF^2}\upsilon^0}
 \simeq~\frac{\tups^{0}-\bn\cdot\tilde{\boldsymbol\upsilon}+z_m(\ta^{0}-\bn\cdot\bta)}{\sqrt{1-z_m^4\cF^2}\upsilon^0},
 \end{equation}
 and then use this to compute the radiated power per unit emission \emph{proper} time,
 \begin{equation}\label{Prad}
 P_{\mbox{\scriptsize far}}\equiv-\frac{dE}{d\tau_r}=\lim_{R\to\infty} \frac{R^2}{\sqrt{1-z_m^4\cF^2}}
 \int d\Omega\,\left[\tups^{0}-\bn\cdot\tilde{\boldsymbol\upsilon}+z_m(\ta^{0}-\bn\cdot\bta)\right]\cE_{\mbox{\scriptsize rad}}~.
 \end{equation}

 Plugging (\ref{e1rad}) into this master formula, and carrying out the angular integration with the help of the formulas in Appendix A, we find that the contribution to the radiated power of the quark arising from the terms with no derivatives is given by
 \begin{equation}\label{P1rad}
 P^{(1)}_{\mbox{\scriptsize far}}=\frac{\sqrt{\lambda}}{2\pi}\frac{\tups^0\ta^2}{\sqrt{1-z_m^4\cF^2}}~.
 \end{equation}
 As a check, notice that via (\ref{tau2}) this translates into $-dE^{(1)}/d\ti_r=\sqrt{\lambda}\ta^2/2\pi$, which in the limit of infinite quark mass correctly agrees with the corresponding result of \cite{HIMT}: it is simply the Lienard rate, first obtained in \cite{mikhailov} through a worldsheet analysis. For finite mass, $z_m>0$, (\ref{P1rad}) differs from the Lienard rate. When rewritten directly in terms of the physical variables by means of (\ref{vtilde}) and (\ref{atilde}), it reads
  \begin{equation}\label{P1rad2}
 P^{(1)}_{\mbox{\scriptsize far}}=\frac{\sqrt{\lambda}}{2\pi}z_m^2\cF^2\left(\frac{v^0-z_m^2\bF\cdot\bv}{1-z_m^4\cF^2}\right)~.
 \end{equation}
Interestingly, this agrees in full with the rate of energy loss (\ref{radiationrate}), derived in \cite{lorentzdirac,damping} by generalizing the worldsheet calculation of \cite{mikhailov}.

 Let us now consider the contributions to the energy density that do have $\tau$-derivatives. Plugging (\ref{largeR}) into (\ref{e12}) we obtain
 \begin{eqnarray}\label{e2rad}
&&\!\!\!\!\cE^{(2)}_{\mbox{\scriptsize far}}=\frac{\sqrt{\lambda}}{8\pi^2}\frac{1}{R^2(\tv^0-\bn\cdot\btv+z_m(\ta^0-\bn\cdot\bta))} \p_{\ttau_r}\Bigg[\frac{z_m\ta^2+\tv^0(\bn\cdot\bta)-\ta^0(\bn\cdot\btv)}{(\tv^0-\bn\cdot\btv)^2[\tv^0-\bn\cdot\btv+z_m(\ta^0-\bn\cdot\bta)]} \nonumber \\
&&\qquad\qquad\qquad\qquad\qquad
-\frac{(1+z_m^2\ta^2)[\btv\cdot\bta-(\bn\cdot\btv)(\bn\cdot\bta)]+z_m^2(\ta\cdot \tj)[\btv^2-(\bn\cdot\btv)^2]}{(\tv^0-\bn\cdot\btv)[\tv^0-\bn\cdot\btv+z_m(\ta^0-\bn\cdot\bta)]^2}
\nonumber \\
&&\qquad\qquad\qquad\qquad\qquad
-\frac{z_m[\bta^2-(\bn\cdot\bta)^2+\btv\cdot\btj-(\bn\cdot\btv)(\bn\cdot\bta)]}{(\tv^0-\bn\cdot\btv)[\tv^0-\bn\cdot\btv+z_m(\ta^0-\bn\cdot\bta)]^2}
\\
&&\qquad\qquad\qquad\qquad\qquad
+\frac{ \left[(\tv^0-\bn\cdot\btv)[\ta^0-\bn\cdot\bta+\frac{z_m}{2}(\tj^0-\bn\cdot\btj)]+\frac{z_m}{2}(\ta^0-\bn\cdot\bta)^2\right]
}{(\tv^0-\bn\cdot\btv)^2[\tv^0-\bn\cdot\btv+z_m(\ta^0-\bn\cdot\bta)]^3}
\nonumber\\.
&&\qquad\qquad\qquad\qquad\qquad
\times\left\{\frac43+2z_m[\btv\cdot\bta-(\bn\cdot\btv)(\bn\cdot\bta)] +(1+z_m^2\ta^2)(\btv^2-(\bn\cdot\btv)^2)\right\} \Bigg]~.
\nonumber
\end{eqnarray}
 If we keep the full $z_m$-dependence, the denominators in this expression make it difficult to carry out the angular integrals needed to obtain the radiated power $P^{(2)}_{\mbox{\scriptsize rad}}$. We will therefore settle for examining it in the regime of non-violent motion where $z_m\ta\ll 1$, retaining only terms up to next-to-leading order in this small quantity. Running through the same procedure as before, and using the table of angular integrals in Appendix A, one ultimately arrives at
 \begin{eqnarray}\label{P2rad}
 P^{(2)}_{\mbox{\scriptsize far}}&=&
 \frac{\sqrt{\lambda}}{2\pi}
 \p_{\ttau_r}\left[-\frac{1}{9}\ta^0
 +z_m\ta^2\tups^0\left(\frac{17}{18}+\frac{1}{4|\tilde{\boldsymbol\upsilon}|^2}\right)
 +z_m(\ta^0)^2\tups^0\frac{3}{4|\tilde{\boldsymbol\upsilon}|^4}
-z_m \tj^0\left(\frac{19}{18}+\frac{1}{2|\tilde{\boldsymbol\upsilon}|^2}\right)
 \right.\nonumber\\
{}&{}&\qquad\qquad\qquad\left.
-z_m\left((\ta^0)^2\left(\frac{1}{4|\tilde{\boldsymbol\upsilon}|^3}+\frac{3}{4|\tilde{\boldsymbol\upsilon}|^5}\right)
 +\frac{\bta^2}{4|\tilde{\boldsymbol\upsilon}|^3}-\frac{\tups^0 \tj^0}{2|\tilde{\boldsymbol\upsilon}|^3}\right)\mbox{arctanh}\frac{|\tilde{\boldsymbol\upsilon}|}{\tups^0}\right]~.
 \end{eqnarray}

 The first term in this expression agrees with the result obtained in the infinite-mass calculation of \cite{HIMT}. We learn then that the $z_m>0$ contributions enter only at the next order.
 As usual, these include not only the  $z_m$-dependence shown explicitly but also the one that arises when (\ref{tau2})-(\ref{atilde}) are used to translate from the auxiliary to the physical variables.

 Since we have already found that $P^{(1)}_{\mbox{\scriptsize far}}$ reproduces (to all orders in $z_m$) the modified Lienard rate of radiation (\ref{radiationrate}) known from the worldsheet analysis, we confirm the suspicion of \cite{HIMT}: the above non-vanishing result for $P^{(2)}_{\mbox{\scriptsize far}}$ can only be an intrinsic-energy contribution.\footnote{One might wonder if somehow it is in the worldsheet approach where the radiative and intrinsic components have not been properly identified. We discuss the potential ambiguity in Appendix B, and show that it cannot change our conclusions.} Some of the specific properties of this expression that would be confusing if we were to insist on interpreting it as radiation were already pointed out in \cite{HIMT} (see also \cite{baier}), including the fact that it is a total derivative, and therefore gives no net integrated contribution in the cases where the quark moves periodically or is accelerated only during a finite interval of time.
 This is closely related to what we regard as the clearest giveaway of $P^{(2)}_{\mbox{\scriptsize far}}$:
 whereas (\ref{P1rad}) is always positive, (\ref{P2rad}) can be either positive or negative. In other words, $P^{(1)}_{\mbox{\scriptsize far}}$ always represents energy flow away from the quark, which is what we expect for radiation in our purely outoing setup,\footnote{Recall that Mikhailov's embedding (\ref{MikSol}) or (\ref{mikhsolzm}) is purely retarded, and is thereby dual to a gluonic field configuration that is purely outgoing.} but $P^{(2)}_{\mbox{\scriptsize far}}$ tells us that, in certain circumstances, energy can flow from infinity toward the quark. This is just what we expect for the intrinsic component of the gluonic field: it is still adhered to the quark, so, with an appropriate tug, it should be possible to pull it in. What is remarkable, of course, is that the $1/R^2$ falloff of this intrinsic tail means that we can pull it in even from arbitrarily far away.

 The authors of \cite{HIMT} noted that the form of the first term of (\ref{P2rad}) is compatible with
 the order $z_m^0$ term of the intrinsic energy obtained from the worldsheet analysis, as can be verified by using (\ref{force}) in (\ref{pq}).
 The numerical coefficient is different, which is natural because we expect most of the intrinsic energy \emph{not} to be contained in the $1/R^2$ tail. For the same reason, we should not be troubled by the fact that the structure seen in the order $z_m$ terms does not resemble the corresponding terms of the total intrinsic energy. But there is one issue that we \emph{should} be troubled by: knowing that $\cE^{(2)}_{\mbox{\scriptsize far}}$ encodes the intrinsic and not the radiative component of the gluonic field, even at large $R$ it is not expected  to conform to
 the relation $\expec{T^{i0}_{\mbox{\scriptsize far}}}\simeq n^i \expec{T^{00}_{\mbox{\scriptsize far}}}$ that we used in (\ref{dEdt}) to turn our result for the energy density into an energy flux. This is not to say that the presence of an intrinsic component in the far zone is in doubt (if there were only radiation, then (\ref{P2rad}) would be reliable), but only that to determine its contribution to the radiated power one must carry out a direct computation of the energy flux sourced by the quark. We suspect that this is also behind the less-than-transparent behavior of most of the order $z_m$ terms in (\ref{P2rad}) under boosts. We leave for future work the task of working out the full energy-momentum tensor sourced by the quark \cite{leon}.

In spite of these limitations, there is one non-trivial check that we can perform on our expression for the intrinsic power. The most worrisome feature of (\ref{P2rad}) is the appearance of $|\tilde{\boldsymbol\upsilon}|$ in the denominators of most of the terms, which could cause a divergence in the limit of small velocities. But, using
  $\mbox{arctanh}(|{\boldsymbol\upsilon}|/\tups^0)\simeq|\tilde{\boldsymbol\upsilon}|/\tups^0+(|\tilde{\boldsymbol\upsilon}|/\tups^0)^3/3+
 (|\tilde{\boldsymbol\upsilon}|/\tups^0)^5/5+\cdots $, we can check that all the dangerous dependence cancels out in this limit, and the expression reduces to
\begin{equation}\label{NR-P2rad2}
    P^{(2)}_{\mbox{\scriptsize far}}\simeq
    \frac{\sqrt{\lambda}}{2\pi}
    \p_{\ttau_r}\left[-\frac{1}{9}\ta^0
    -\frac{53}{45}z_m(\ta^0)^2
    +z_m\bta^2\left(\frac{10}{9}+\frac{19|\btv|^2}{45}\right)
   -z_m \tj^0\left(\frac{8}{9}+\frac{11|\btv|^2}{60}\right)\right]~,
     \end{equation}
    where we have kept only terms up to second order in $|\btv|$ (notice $(\ta_0)^2$ is internally already of this order).
    The leading low-velocity terms here, when rewritten in terms of the physical three-velocity $\vec{\upsilon}$, three-acceleration $\vec{a}$ and three-jerk $\vec{j}$ read
           \begin{equation}\label{NR-P2rad3}
                P^{(2)}_{\mbox{\scriptsize far}}=
                \frac{\sqrt{\lambda}}{2\pi}
                \p_{t_r}\left[-\frac{1}{9}\vec{\upsilon}\cdot\vec{a}
                +\frac{3}{9}z_m\vec{a}^2
               -\frac{7}{9}z_m\vec{\upsilon}\cdot\vec{j}\right]~,
                \end{equation}
      which has precisely the same structure although not the same coefficients as the non-relativistic limit \cite{damping} of the intrinsic energy (\ref{pq}),
      \begin{eqnarray}
      E(t)
      %&=&\frac12 m\left(\sum_{l=1}^\infty (-z_m)^{l-1}\frac{d^l\vec{x}}{dt^l}\right)^2\nonumber \\
      &=&\frac12 m \vec{\upsilon}^2
      +\frac{\sqrt{\lambda}}{2\pi} \left(-\vec{v}\cdot\vec{a}+\frac12 z_m \vec{a}^2 +z_m \vec{\upsilon}\cdot\vec{j}\right)~.
      \end{eqnarray}

 The existence of a long-distance tail in the intrinsic component of the CFT fields is closely related to an observation made very recently by Lewkowycz and Maldacena \cite{lewkomalda}.\footnote{We thank Aitor Lewkowycz for discussions on this relation.} In the context of an exploration of various quantities associated with infinitely massive quarks (i.e., Wilson loops) that can be computed exactly via localization (see also \cite{correa,fiol} and references therein), these authors noticed a discrepancy between the coefficient in the one-point function of the energy-momentum tensor sourced by the quark and the so-called Bremsstrahlung function that determines, among other things, the quark's rate of radiation.
The specific motion under scrutiny in \cite{lewkomalda} was uniform acceleration, for which even in classical electromagnetism there are well-known difficulties in separating the intrinsic and radiative components of the energy. Lewkowycz and Maldacena showed that an invariant subtraction procedure based on the fact that the quark also sources the dimension 2 scalar operator $O_2\sim (\Phi_1)^2+\ldots$ (where the subindex 1 denotes the direction along which the string endpoint is oriented on the $S^5$) successfully removes the discrepancy. In other words, at least in the case of uniform acceleration, the portion that is subtracted is precisely the intrinsic contribution.

  When $z_m=0$, it is easy to check that for a static quark this subtraction correctly removes the entire profile (\ref{edstatic}), which is of course purely intrinsic. For the uniformly accelerated case, it removes \cite{aitor}
   nothing more and nothing less than the value of (\ref{P2rad}) obtained previously in \cite{HIMT} (which was duly noted to be unreliable for this type of motion, which is not bounded).
  It seems then that, at least for infinite quark mass, the procedure conjectured in \cite{lewkomalda} is indeed the appropriate way to separate the intrinsic component of the CFT fields sourced by the quark. {}From the connection with the results of \cite{HIMT} and the present paper, we learn that this is not just an issue specific to the case of uniform acceleration examined in \cite{lewkomalda}, but is in fact due to the presence of a $1/R^2$ tail of the intrinsic energy for arbitrary (nonstationary) trajectories. In the opposite direction, the findings of \cite{lewkomalda} teach us that this tail is entirely due to the fact that our quark excites a profile in the (conformally coupled) scalar fields of the gauge theory.

  It is important to emphasize that, while we have found convincing evidence that $\cE^{(1)}$ and $\cE^{(2)}$ respectively encode the radiation and intrinsic components of the gluonic field when examined at large distances for fixed emission time $t_r$, this \emph{does not} imply that the same identification holds when these 2 contributions of (\ref{e12}) are examined at arbitrary spacetime locations. In fact, as we mentioned in Section \ref{totalsec}, we have explicitly checked that the integral of $\cE^{(1)}$ at fixed observation time $t$ differs from the accumulated energy loss by radiation. We have thus been unable, at finite quark mass, to find a local way to separate the energy density into intrinsic and radiative components by direct calculation, in analogy with \cite{teitelboim}. A very interesting question then, which we leave for future work, is whether the indirect procedure of \cite{lewkomalda}, which brings into play the scalar expectation value $\expec{O_2}$, succeeds in providing us with this separation for arbitrary motion, even when $z_m>0$. Notice this is not at all guaranteed. Already in the case of infinite mass, we know from the perturbative calculations in \cite{liusynchrotron,HIMT} that at least at weak coupling the scalar fields do make a significant contribution to the genuine radiation, which the prescription of \cite{lewkomalda} must then avoid subtracting.

\section{Distant vs. Small and a Dynamical UV/IR Connection} \label{uvirsec}

In this final section, we would like to emphasize another surprising property of the energy density (\ref{e12}).
As explained in the Introduction, it is natural to expect
 the finite size $z_m$ of the quark  not to have a noticeable effect on the gluonic profile at distances $R\gg z_m$, both from the
 gauge theory perspective and from
the usual UV/IR intuition \cite{uvir} in the gravity description.
With the long-distance results of Section \ref{farsec} in hand, we can now put this expectation to the test, in the hope of learning more not only about radiation in strongly-coupled CFTs, but also about the UV/IR map itself.

We have found that the energy density (\ref{e12}) of a quark with finite size simplifies, in the long-distance limit $z_m/R\to 0$, to (\ref{e1rad})$+$(\ref{e2rad}). We see manifestly that this does not agree with the result for a pointlike quark \cite{HIMT}, obtained by setting $z_m=0$ in (\ref{e12}): not only are there terms in (\ref{e1rad}) and (\ref{e2rad}) that  explicitly contain $z_m$, but also there is hidden dependence on the quark Compton radius that is revealed upon translating from the tilde to the non-tilde variables via (\ref{tau2})-(\ref{atilde}). Moreover, the two expressions depend on the quark variables evaluated at different instants, because the conditions (\ref{tautilderet}) or (\ref{tauret}), which respectively determine the tilde or nontilde versions of the retarded time, depend on $z_m$. As we saw in Section \ref{totalsec}, for finite $z_m$ the physical emission event $x(\tau_r)$ lies on a hyperboloid a proper time $z_m$ to the past of the observation event $x$, while for $z_m=0$ it would instead lie on the past lightcone of $x$. The two instants are not simply related, even if we measure the field at large distances from the quark. Indeed, it follows from the definitions that, for large $R$, the corresponding retarded times differ by $\Delta t\sim z_m\tilde{\gamma}$, and in this interval the quark motion can in general change
appreciably.\footnote{The origin of this time difference is readily understood in the bulk picture. Given the structure of the embedding (\ref{MikSol}), we know that the full string at some time $t$ encodes the gluonic field sourced by the quark at all times up to $t$. But when we truncate the string at $z_m$, the time delay associated with propagation from the AdS boundary to this radial depth implies that the string only knows about the behavior of the endpoint at $z=0$ up to a time that is earlier by precisely $\Delta t$.  Naively, we might expect the missing information to become irrelevant when we integrate over the contributions of all string bits and examine the result from a distant observation point, but we know from \cite{HIMT,trfsq} and the present paper that this integral actually gives rise to just a surface term, which is then evaluated precisely at the last accessible emission point. So, even just from this fact alone, the initial expectation that the pointlike and nonpointlike profiles would directly match is too much to ask. The best we could hope is that the 2 expressions would match if we by hand shift the time argument appropriately before making the comparison at large distances. But, on the other hand, performing the comparison this way is certainly peculiar from the gravity side of the correspondence, because it means that we compare 2 rather different string profiles. This serves to underscore how extraordinary it is  that the integral over the entire string boils down to just an endpoint contribution, as established in \cite{HIMT,trfsq}.}

 Notice that the mismatch due to these various surviving dependences on $z_m$ occurs both for (\ref{e1rad}) and (\ref{e2rad}), which as discussed in Section \ref{farsec} yield
contributions to the power that correspond respectively to the radiation and intrinsic components of the gluonic field. In fact, the same type of mismatch is found in the
purely intrinsic-field observable $\expec{\tr F^2(x)}$. The full result for finite quark mass is \cite{trfsq}
\begin{eqnarray}\label{trfsqfull}
\expec{\trFsq(x)}&=&{\sqrt{\lambda}\over 32\pi^{2}}
\frac{1}{\left((x-\tx)\cdot\tv\right)^2 \left[z_m+(x-\tx)\cdot(\tv+\ta z_m)\right]^5}\\
{}&{}& \times\bigg\{2\left((x-\tx)\cdot\tv\right)^3
             +z_m^3 \left(1+(x-\tx)\cdot\ta\right)^3
\nonumber\\
&&\;\;\quad +z_m\left((x-\tx)\cdot\tv\right)^2\left[2+(x-\tx)\cdot(2\ta-4\tj z_m-\ts z_m^2)-\ta^2 z_m^2\right]
\nonumber\\
{}&{}&\;\;\quad +z_m^2(x-\tx)\cdot\tv\left[4+4\left((x-\tx)\cdot\ta\right)^2
+2(x-\tx)\cdot\tj z_m+3\left((x-\tx)\cdot\tj\right)^2 z_m^2 \right.
\nonumber\\
&&\;\;\quad -(x-\tx)\cdot\ts z_m^2-\ta^2z_m^2
   +%\left.\left.
   (x-\tx)\cdot\ta\left(8+(x-\tx)\cdot(2\tj-\ts z_m)z_m-\ta^2 z_m^2\right)\Big]\bigg\}~,
\nonumber
\end{eqnarray}
which reduces at long distances to
\begin{eqnarray}
\label{trfsqfar}
\expec{\trFsq(x)}_{\mbox{\scriptsize far}}&=&{\sqrt{\lambda}\over 16\pi^{2}}
\frac{1}{\left[(x-\tx)\cdot\tv\right]^4} \frac{1}{ \left[1+z_m\left[(x-\tx)\cdot\ta/(x-\tx)\cdot\tv\right]\right]^5}\\
{}&{}&\quad \times\left\{ 1+z_m\left[\frac{(x-\tx)\cdot\ta}{(x-\tx)\cdot\tv}\right]
+2z^2_m\left[\frac{(x-\tx)\cdot\ta}{(x-\tx)\cdot\tv}\right]^2 - 2z^2_m\left[\frac{(x-\tx)\cdot\tj}{(x-\tx)\cdot\tv}\right]\right.
\nonumber \\
{}&{}&\quad\quad\;+\frac{z^3_m}{2}\left[\frac{(x-\tx)\cdot\ta}{(x-\tx)\cdot\tv}\right]^3
-\frac{z^3_m}{2}\left[\frac{(x-\tx)\cdot\ts}{(x-\tx)\cdot\tv}\right]
+z^3_m\left[\frac{(x-\tx)\cdot\ta}{(x-\tx)\cdot\tv}\right]\left[\frac{(x-\tx)\cdot\tj}{(x-\tx)\cdot\tv}\right]
\nonumber\\
{}&{}&\quad\quad\;+\frac{3}{2}z^4_m\left[\frac{(x-\tx)\cdot\tj}{(x-\tx)\cdot\tv}\right]^2
-\frac{1}{2}z^4_m\left[\frac{(x-\tx)\cdot\ta}{(x-\tx)\cdot\tv}\right]\left[\frac{(x-\tx)\cdot\ts}{(x-\tx)\cdot\tv}\right]\bigg\}~,
   \nonumber
\end{eqnarray}
and so evidently does not coincide with the pointlike limit of (\ref{trfsqfull}), which is simply
\begin{equation}
\label{trfsqpoint}
\expec{\trFsq(x)}_{\mbox{\scriptsize point}}={\sqrt{\lambda}\over 16\pi^{2}}
\frac{1}{\left[(x-\tx)\cdot \tv \right]^4}~.
\end{equation}

We thus learn that, under appropriate conditions, even very far from the quark it would be possible to detect whether or not it has a finite size, through a measurement of the $1/R^4$ tail of $\expec{\trFsq(x)}$, or, even more remarkably, of the longer-range $1/R^2$ behavior of $\expec{T_{00}(x)}$.

By inspection of (\ref{trfsqfar}) and (\ref{trfsqpoint}) we see that a match is only achieved for a generic (distant) observation point if $z_m$ is small enough that
%$|(x-\tx)\cdot\ta| z_m, |(x-\tx)\cdot\tj| z_m^2, |(x-\tx)\cdot\ts| z_m^3\ll |(x-\tx)\cdot\tv|$.
\begin{equation} \label{nonviolent}
\left|\frac{(x-\tx)\cdot\ta z_m}{(x-\tx)\cdot\tv}\right|~,
 \left|\frac{(x-\tx)\cdot\tj z_m^2}{(x-\tx)\cdot\tv}\right|~,
\left|\frac{(x-\tx)\cdot\ts z_m^3}{(x-\tx)\cdot\tv}\right|\ll 1~.
\end{equation}
After canceling out the common factor $t-t(\tau_r)\simeq R$, the right-hand side of these inequalities becomes $\tilde{\gamma}(1-|\vec{\tv}|\cos\nu)$, where $\nu$ is the angle between $\btv$ and $\bn$. For generic values of $\nu$, this scales like $\tilde{\gamma}$, so we require
that\footnote{In the special case $\nu=0$ the right-hand side attains its smallest possible value, $\tilde{\gamma}(1-|\vec{\tv}|)\simeq 2/\bar{\gamma}$, and we are led instead to the requirement that
$\tilde{\gamma} z_m|\ta^{\mu}|, \tilde{\gamma}z^2_m|\tj^{\mu}|, \tilde{\gamma} z^3_m|\ts^{\mu}|\ll 1$.}
\begin{equation} \label{nonviolent2}
z_m\frac{|\ta^{\mu}|}{\tilde{\gamma}},
z^2_m\frac{|\tj^{\mu}|}{\tilde{\gamma}},
z^3_m\frac{|\ts^{\mu}|}{\tilde{\gamma}}\ll 1~.
\end{equation}
%$z_m|\ta^{\mu}/\tilde{\gamma}|, z^2_m|\tj^{\mu}|/\tilde{\gamma}, z^3_m|\ts^{\mu}|/\tilde{\gamma}\ll 1$.
 This guarantees that in (\ref{trfsqfar}) the factor raised to the fifth power in the denominator and the long expression within braces reduce to unity. These conditions also ensure that the variables describing the state of motion of the quark do not change appreciably in the interval
 $\Delta t$, implying that the discrepancy in retarded times mentioned above becomes negligible. And examining (\ref{e1rad}) and (\ref{e2rad}) we see that exactly the same conditions are needed for $\expec{T_{00}(x)}$ in the far zone to reduce to the pointlike result.
With the aid of (\ref{atilde}), we find that in terms of nontilde variables the requirement is that
$z_m^2\bar{\cF}^{\mu}, z_m^3 \dot{\bar{\cF}}^{\mu},  z_m^4\ddot{\bar{\cF}}^{\mu}\ll 1$,
which stipulates in particular that the external force be much smaller than its critical value (\ref{Fcrit}). Equivalently, through the equation of motion (\ref{eom}), the conditions read
\begin{equation} \label{nonviolent3}
z_m|a^{\mu}|,
z^2_m |j^{\mu}|,
z^3_m |s^{\mu}|\ll 1~.
\end{equation}

Purely at the level of dimensional analysis, it is clear that the reason why $z_m/R\to 0$ does not lead to the same results as $z_m=0$ is that the temporal variation in the motion brings another scale to the problem. The physics behind this, in gauge theory language, is that the quark is a dynamical blob, that gets noticeably distorted when its motion is sufficiently violent. It is not obvious (at least to us)
 that this distortion should be appreciable even in the far zone, but in retrospect this is in fact necessary for consistency with the worldsheet analysis of \cite{mikhailov,dragtime,damping},
 precisely when the quark radiation rate (\ref{radiationrate}) and dispersion relation (\ref{pq}) differ significantly from the standard pointlike expressions.

 {}From the gravity perspective, it should \emph{a priori} not be too surprising that the presence or absence of the $z\le z_m$ portion of the string could make a difference even far away, because this piece actually has infinite proper length and infinite energy. Still, naive UV/IR reasoning in the context of AdS would lead us to believe that this UV information about the string should be negligible at large enough distances. We have found instead that the mapping between the gravity and gauge theories is sufficiently nonlocal that, unless we impose conditions (\ref{nonviolent3}), bulk data at radial positions $\le z_m$ in AdS can have an important influence on the gluonic profile even at distances $R\gg z_m$ away from the quark. We are thus learning that, in a dynamical setting, the UV/IR connection assumes a more nuanced form: we must keep track not only of the distance, but also of our \emph{temporal resolution}. Indeed, when interpreted in this light it seems natural that, if we forfeit information about the UV portion of the string, we lose the ability of registering the high-frequency components of the motion of the dual quark.

It is worth noting that what conditions (\ref{nonviolent3}) directly forbid is the presence of high-frequency components in the motion of the quark, and the corresponding cutoff in the frequency of the gluonic field itself can be different. Indeed, we see from (\ref{dtdtr}) that there is a relative factor of $1-\bn\cdot\boldsymbol\upsilon$ between the frequencies defined with the retarded time for the source and the observation time. This is precisely the factor that is responsible for beaming of the gluonic field when the motion is ultrarelativistic \cite{liusynchrotron,veronika,beaming}, so the cutoff in observed frequencies is proportionately larger than the quark frequencies for this type of motion. This seems consistent with the fact that, whereas the string is truncated at $z=z_m$ and is thus truly missing its near-boundary component, we do read off the behavior of the bulk fields in the $z\to 0$ region, so we are not directly enforcing a UV cutoff on the CFT fields themselves.

For the case of uniform circular motion, we show in Appendix C that the restriction (\ref{nonviolent2}) to nonviolent motion is in some tension with the requirement that the string bits move relativistically, which is the basis of Hubeny's shock wave proposal \cite{veronika}, described in the Introduction. This means that we are unable to fully understand the success of her approximation scheme in this case. More generally, when attempting to apply the method of \cite{veronika} to other types of quark trajectories, condition  (\ref{nonviolent2}) should be taken into account, in combination with the requirements previously identified in \cite{beaming}.

\section*{Acknowledgements}
We are grateful to Mariano Chernicoff, Veronika Hubeny and Aitor Lewkowycz for useful discussions.
The research of CA is supported in part by the Department of Energy via Award DE-SC0009987.
AG is partially supported by Mexico's National Council of Science and Technology (CONACyT) grant 104649, DGAPA-UNAM grant IN110312, as well as sabbatical fellowships from DGAPA-UNAM and CONACyT. He would also like to thank the Department of Physics of Princeton University, and Igor Klebanov in particular, for hosting his sabbatical.
JFP is partially supported by the National Science Foundation under Grant No. PHY-1316033 and by the Texas Cosmology Center.

\section*{Appendix A: Angular Integrals}
In the evaluation of the angular integral of the part of the energy density involving $\tau$ derivatives, we find the  expression
\begin{eqnarray}
P^{(2)}_{\textrm{rad}}&=&\frac{z_m\sqrt{\lambda}}{8 \pi^2}\partial_{\tilde{\tau}_r}\bigg[
4\pi \tv^0 \ta^2-\ta^0\tv^0\int \frac{\dif \Omega\,(\bn\cdot\bta)}{(\tv^0-\bn\cdot\btv)^4}+\tv^0\int \frac{\dif \Omega\, (\bn\cdot\bta)^2}{(\tv^0-\bn\cdot\btv)^4}+(\ta^0)^2\int \frac{\dif \Omega\, (\bn\cdot\btv)}{(\tv^0-\bn\cdot\btv)^4}\nonumber \\
&& \qquad \qquad
+\int \frac{\dif \Omega\, (\bn\cdot\btv)(\bn\cdot\btj)}{(\tv^0-\bn\cdot\btv)^3}+\frac{2}{3}\int\frac{\dif\Omega\,(\tj^{0}-\bn\cdot\btj )}{(\tv^0-\bn\cdot\btv)^4}+\frac{1}{2}\int\frac{\dif\Omega\, (\bn\times \btv)^2(\tj^{0}-\bn\cdot\btj )}{(\tv^0-\bn\cdot\btv)^4}
\nonumber \\
&& \qquad \qquad
+4\int \frac{\dif \Omega\, (\bn\cdot\btv)(\bn\cdot\bta)^2}{(\tv^0-\bn\cdot\btv)^4}-(\bta^2+\btv\cdot\btj)\int \frac{\dif \Omega\, }{(\tv^0-\bn\cdot\btv)^3} +\int \frac{\dif \Omega\, (\bn\cdot\bta)^2}{(\tv^0-\bn\cdot\btv)^3}
\nonumber \\
&& \qquad \qquad
-\frac{10}{3}\int \frac{\dif \Omega\, (\ta^{0}-\bn\cdot\bta )^2}{(\tv^0-\bn\cdot\btv)^5}-\frac{5}{2}\int\frac{\dif\Omega\, (\bn\times \btv)^2(\ta^{0}-\bn\cdot\bta )^2}{(\tv^0-\bn\cdot\btv)^5}
\nonumber \\
&& \qquad \qquad
+4(\btv\cdot\bta)\int \frac{\dif \Omega\, (\ta^{0}-\bn\cdot\bta )}{(\tv^0-\bn\cdot\btv)^4}-5\ta^0\int \frac{\dif \Omega\, (\bn\cdot\btv)(\bn\cdot\bta)}{(\tv^0-\bn\cdot\btv)^4}
\bigg]
\end{eqnarray}
which involves a number of basic integrals just at first order in $z_m$. Many of these integrals can be derived from a subset of them by taking partial derivatives with respect to the parameter $\tv^0$, to increase the power of the common denominator $(\tv^0-\bn\cdot\btv)$. However, for completeness we will present here the simplified results for each of the integrals appearing in the previous expression:
\begin{eqnarray}
\label{Table of Integrals}
\int \frac{\dif\Omega}{(\tv^0-\bn\cdot\btv)^3}&=&4\pi\tv^0
\nonumber\\
\int \frac{\dif\Omega}{(\tv^0-\bn\cdot\btv)^4}&=&\frac{4\pi}{3}(|\btv|^2+3(\tv^0)^2)
\nonumber\\
\int \frac{\dif \Omega (\bn\cdot\btv)}{(\tv^0-\bn\cdot\btv)^4}&=&\frac{16\pi}{3}\tv^0 |\btv|^2
\nonumber\\
\int \frac{\dif \Omega (\bn\cdot\bta)}{(\tv^0-\bn\cdot\btv)^4}&=&\frac{16\pi}{3}\tv^0(\btv\cdot\bta)
\nonumber\\
\int \frac{\dif \Omega\, (\bn\cdot\bta)^2}{(\tv^0-\bn\cdot\btv)^3}&=&\frac{\pi}{|\btv|^5}\bigg[2|\btv|^3|\bta|^2\tv^0 +2(\tv^0)^2(\ta^0)^2(2|\btv|(\tv^0)^3-5|\btv|\tv^0)
\nonumber \\
&&\qquad \qquad+(|\btv|^2|\bta|^2-3(\tv^0)^2(\ta^0)^2)\ln\left(\frac{\tv^0-|\btv|}{\tv^0+|\btv|}\right) \bigg]
\nonumber\\
\int \frac{\dif \Omega (\bn\cdot\bta)^2}{(\tv^0-\bn\cdot\btv)^4}&=&\frac{4\pi}{3}(|\bta|^2+4(\ta^0)^2(\tv^0)^2)
\nonumber\\
\int \frac{\dif \Omega\, (\ta^{0}-\bn\cdot\bta )^2}{(\tv^0-\bn\cdot\btv)^5}&=&\frac{4\pi}{3}\tv^0 \ta^2
\nonumber
\end{eqnarray}
\begin{eqnarray}
\int \frac{\dif \Omega\, (\tj^0-\bn\cdot\btj)}{(\tv^0-\bn\cdot\btv)^4}&=&-\frac{4\pi}{3}\tj^0+\frac{16\pi}{3}\tv^0 \ta^2
\nonumber\\
\int \frac{\dif \Omega\, (\bn\cdot\btv)(\bn\cdot\bta)}{(\tv^0-\bn\cdot\btv)^4}&=&\frac{4\pi}{3}(\btv\cdot\bta)(3|\btv|^2+(\tv^0)^2)
\nonumber\\
\int \frac{\dif \Omega\,(\bn\cdot\btv) (\bn\cdot\btj)}{(\tv^0-\bn\cdot\btv)^3}&=&
\frac{2\pi(\btv\cdot\btj)}{|\btv|^3}\Big[4|\btv|^3\tv^0-2|\btv|(\tv^0)^3-\ln\left(\frac{\tv^0-|\btv|}{\tv^0+|\btv|}\right) \Big]
\nonumber\\
\int \frac{\dif \Omega\, (\bn\cdot\btv)(\bn\cdot\bta)^2}{(\tv^0-\bn\cdot\btv)^4}&=&\frac{\pi}{3|\btv|^5}\bigg[-3(|\btv|^2|\bta|^2-3(\tv^0)^2(\ta^0)^2)\ln\left(\frac{\tv^0-|\btv|}{\tv^0+|\btv|}\right)
\nonumber\\
&&\qquad \qquad +2|\btv|(\tv^0)^3(\ta^0)^2(1+8(\tv^0)^4-22|\btv|^2) \nonumber\\
&&\qquad \qquad -2|\btv|^3|\bta|^2\tv^0(3-2|\btv|^2) \bigg]
\nonumber\\
\int\frac{\dif\Omega\, (\bn\times \btv)^2(\tj^{0}-\bn\cdot\btj )}{(\tv^0-\bn\cdot\btv)^4}&=&\frac{\pi}{|\btv|^3}\bigg[2(\btv\cdot\btj)\ln\left(\frac{\tv^0-|\btv|}{\tv^0+|\btv|}\right)+\frac 83|\btv|^5j^0
\nonumber\\
&&\qquad \qquad +(\btv\cdot\btj)\left(-\frac{20}{3}|\btv|^3\tv^0+4|\btv|(\tv^0)^3 \right)\bigg]
\nonumber\\
\int\frac{\dif\Omega\, (\bn\times \btv)^2(\ta^{0}-\bn\cdot\bta )^2}{(\tv^0-\bn\cdot\btv)^5}&=&\frac{\pi}{3|\btv|^5}\bigg[ -3(|\btv|^2|\bta|^2-3(\tv^0)^2(\ta^0)^2)\ln\left(\frac{\tv^0-|\btv|}{\tv^0+|\btv|}\right)
\nonumber \\
&&\qquad \qquad+2|\btv|\tv^0(\ta^0)^2(4+5(\tv^0)^2-2|\btv|^2(\tv^0)^2)
\nonumber \\
&&\qquad \qquad
-2|\btv|^3|\bta|^2\tv^0(3-2|\btv|^2)\bigg]
\nonumber
\end{eqnarray}

\section*{Appendix B: A Different Energy Partition on the Worldsheet?}

 Given our results in the main text, it seems relevant to explore what margin there is to depart from the canonical identification of (\ref{pq}) as the intrinsic four-momentum of the quark, and (\ref{radiationrate}) as its rate of radiation. When splitting the quark equation of motion (\ref{eom}) as in (\ref{eomsplit}), we could choose to add a total derivative $m (de^{\mu}/d\tau)$ to what we recognize as $dp^{\mu}_q/d\tau$, while subtracting it from $dp^{\mu}_{\mbox{\scriptsize rad}}/d\tau$. This would effect the change
 $p^{\mu}_q\to p^{\prime\mu}_q\equiv p^{\mu}_q+m e^{\mu}$. Such a modification is constrained by two requirements: the new intrinsic four-momentum must still satisfy the mass-shell condition, $p^{\prime 2}_q=-m^2$, so
 it ought to be the case that
 \begin{equation}\label{massshell}
 e^2+\frac{2 e\cdot p_q}{m}=0~,
 \end{equation}
 and it must still reduce to the known expression in the pointlike (infinite-mass) limit,
 \begin{equation}\label{pointlike}
 p^{\prime\mu}_q\simeq p^{\mu}_q\simeq m \upsilon^{\mu}\quad \mbox{for} \quad \frac{\sqrt{\lambda}}{2\pi}|\cF^{\mu}|\ll m^2~.
 \end{equation}

  The main question that interests us is whether a modification of this sort could allow the radiation rate $dp^{\prime 0}_{\mbox{\scriptsize rad}}/d\tau$ to match onto the full $1/R^2$ result $P^{(1)}_{\mbox{\scriptsize far}}+P^{(2)}_{\mbox{\scriptsize far}}$, and thus bring us back to the standard situation where the entire gluonic profile in the far zone is identified as radiation. Since we have found that (\ref{P1rad}) already matches (\ref{radiationrate}), the question is whether the new piece $-(\sqrt{\lambda}/2\pi z_m) de^{0}/d\tau$ that we would identify as radiated energy can match (\ref{P2rad}).

 The extra piece $e^{\mu}$ is a four-vector that must be constructed out of the available data, namely the quark velocity $\upsilon^{\mu}$, the external force $\cF^{\mu}$, and possibly the higher time derivatives $d\cF^{\mu}/d\tau$, etc. Through the equation of motion (\ref{eom}), this data set is equivalent to $\upsilon^{\mu}, a^{\mu}, j^{\mu}$, etc. But for comparison against (\ref{P2rad}), it is easier to parametrize $e^{\mu}$ in terms of the auxiliary data. Noticing that $p^{\mu}_q=m\tups^{\mu}$ and using dimensional analysis, we can write
 \begin{equation}\label{emu}
 e^{\mu}=\tups^{\mu}f_1+z_m\ta^{\mu}f_2+z_m^2\tj^{\mu}f_3+\ldots~,
  \end{equation}
  where the $f_i$ are functions of the available dimensionless scalar combinations $z_m^2\ta^2, z_m^3\ta\cdot\tj, z_m^4\tj^2$, etc. Enforcing (\ref{massshell}) and (\ref{pointlike}), we see that the leading modification for small $z_m$ corresponds to taking $f_2=\mbox{constant}$,
  $f_1=f_2^2 z_m^2 \ta^2/2$ and $f_3=f_4=\ldots=0$. It is plain to see that this  does not agree with (\ref{P2rad}), and neither does the next-to-leading extension involving $f_3$. We thus learn that there is no way to adjust the worldsheet identification of intrinsic and radiated energy to be able to interpret $P^{(2)}_{\mbox{\scriptsize far}}$ as radiation.
  Independently of that, the fact remains that for various reasons it would be peculiar to identify as radiation a \emph{total derivative} of some four-vector like (\ref{emu}). Any such term has the very nontrivial property that, for arbitrary trajectories, its integrated contribution would depend only on the instantaneous state of motion of the quark, and not on its cummulative history.

\section*{Appendix C: Dynamical UV/IR in Uniform Circular Motion}

The analysis presented in section \ref{uvirsec} has an important implication for the approximation scheme of the string gravitational backreaction proposed in \cite{veronika}. As mentioned in the Introduction, in this scheme, the graviton field sourced by the string is approximated as a linear superposition of shock wave contributions from individual string bits. If correct, this would be useful not only as a gravitational explanation of the beaming seen in the gluonic profile, but also as an efficient calculational method.

  More specifically, the approximation of \cite{veronika} is argued to hold for the region where the transverse velocity of the string is relativistic. An encouraging fact is that, for a generic trajectory, the transverse velocity of the string tends to increase with $z$, so one might hope that the approximation of \cite{veronika} will be valid in some region $z>z_r$ on the string that is sufficiently far from the boundary.\footnote{In general, the value of $z_r$ depends on the instantaneous state of motion of the quark/endpoint, and there will also be an upper bound on the relativistic region of the string \cite{beaming}.} The usual intuition about the UV/IR connection would then appear to indicate that the contributions coming from $z>z_r$ would reproduce accurately the near-boundary gravitational backreaction of the string in spatial regions sufficiently far away from the string endpoint. In essence, this is tantamount to the expectation that the presence of absence of the string segment at $z<z_r$ does not change the results considerably at points sufficiently far away from the quark position. However, our main conclusion in Section \ref{uvirsec} is that this  turns out to be false, unless the quark motion is such that its kinematic variables do not change appreciably during time intervals of order $\Delta t \sim \gamma z_r$.

This then places a limitation on the possible validity of the scheme of \cite{veronika} for arbitrary quark trajectories, which would add to the various restrictions enumerated in \cite{beaming}. It is interesting to see how this new condition plays out in the remarkable case of uniform circular motion, in which the original proposal was motivated and numerically tested in \cite{veronika}, and analytically justified in \cite{beaming}.
 In this case the jerk and snap can be written in terms of the 3-velocity $\vec{\upsilon}$ and 3-acceleration $\vec{a}$ as
$j^\mu=-\gamma^3\vec{a}^{\,2}/\vec{\upsilon}^{\,2}(0,\vec{\upsilon})$ and  $s^\mu=-\gamma^2\vec{a}^{\,2}a^\mu/\vec{\upsilon}^{\,2}$. These relations allows us to express the ratios appearing in (\ref{nonviolent}) as
\begin{eqnarray}
\label{circularapproxs}
\left[\frac{(x-\tx)\cdot\ta}{(x-\tx)\cdot\tv}\right]&\simeq&-\gamma|\vec{a}|\left[\frac{\cos\phi}{1-|\vec{v}|\cos\psi}\right]\leq - \gamma^3|\vec{a}|~,\nonumber
\\
 \left[\frac{(x-\tx)\cdot\tj}{(x-\tx)\cdot\tv}\right]&\simeq&\frac{\gamma^2\vec{a}^{\,2}}{\vec{v}^{\,2}}
 \left[\frac{|\vec{v}|\cos\psi}{1-|\vec{v}|\cos\psi}\right]\leq+\frac{\gamma^4|\vec{a}|^2}{|\vec{v}|}~,
\\
\left[\frac{(x-\tx)\cdot\ts}{(x-\tx)\cdot\tv}\right]&\simeq&-\frac{\gamma^2\vec{a}^{\,2}}{\vec{v}^{\,2}}
\left[\frac{(x-\tx)\cdot\ta}{(x-\tx)\cdot\tv}\right]\leq \frac{\gamma^5|\vec{a}|^3}{\vec{v}^{\,2}}~.\nonumber
\end{eqnarray}
In the middle and right-hand side, all quantities are understood to refer to data at $z=0$, and to be evaluated at the appropriate retarded time $\tilde{t}_r$, even though we are omitting some of the corresponding tildes for simplicity. The variables  $\psi$ and $\phi$ are the angles between  $\vec{x}-\vec{\tx}$ and the vectors $\vec{a}$  and $\vec{v}$, respectively. The inequalities are obtained by choosing $\cos\psi=1$ and $\cos\phi=1$, with which the expressions in the middle are maximized.

Plugging the upper limits of  (\ref{circularapproxs}) into the conditions (\ref{nonviolent}) needed for the $z<z_r$ portion of the string to be negligible (and writing $z_r$ in place of $z_m$), we obtain polynomials involving products of $\gamma^3z_r|\vec{a}|$ and ${\gamma|\vec{a}|}z_r/{|\vec{\upsilon}|}$. In order that the total gluonic field (\ref{trfsqpoint}) be well approximated by its truncated counterpart (\ref{trfsqfar}), we must therefore require that
\begin{eqnarray}
\label{truncatedconditions}
 \gamma^3z_r|\vec{a}|\ll1 &\quad\leftrightarrow\quad& \frac{z_r}{\rho_0}\ll \frac{1}{\gamma^3|\vec{\upsilon}|^2}~,\nonumber\\
%\label{truncatedconditions2}
 \gamma z_r|\vec{a}|\lesssim |\vec{\upsilon}| &\quad\leftrightarrow\quad& \frac{z_r}{\rho_0} \lesssim \frac{1}{\gamma|\vec{v}|}~,
\end{eqnarray}
%why weak inequality?
where $\rho_0$ denotes the radius of the circle on which the quark moves.
It is easy to see that these two conditions are not independent: in the case of a relativistic quark, the first implies the second, while in the nonrelativistic case, the opposite is true.
In either case, the criterion is very restrictive. When the quark rotates slowly, (\ref{truncatedconditions}) is incompatible with the condition $\gamma^3|\vec{a}|z_r\ge 1$
required for the points on the string to be relativistic \cite{veronika}. When $\gamma|\vec{v}|\ge 1$, there is no incompatibility because the entire string is relativistic, but (\ref{truncatedconditions}) still places an upper bound on the region of the string that is amenable to the approximation of \cite{veronika}. For other types of quark trajectories, the general conditions (\ref{nonviolent}) or (\ref{nonviolent2}) should similarly be taken into consideration jointly with the other requirements discussed in \cite{veronika,beaming}.

\end{document}